\documentclass[conference]{IEEEtran}

\ifCLASSINFOpdf
  \usepackage[pdftex]{graphicx}
  \DeclareGraphicsExtensions{.pdf,.jpeg,.png}
  \graphicspath{{figures/}}
\else
\fi

\usepackage{subcaption}[tight]
\usepackage{fancyvrb}
\usepackage{xspace}
\usepackage{enumitem}
\usepackage{algpseudocode}
\usepackage{algorithm}
\usepackage{cleveref}
\usepackage{pifont}
\usepackage{tablefootnote}
\usepackage{multirow}
\usepackage[nolist,nohyperlinks]{acronym}
\usepackage[table]{xcolor}

\newcommand{\ie}{i.e., \@}
\newcommand{\eg}{e.g., \@}

\newcommand{\etal}{et al.\xspace}

\newcommand{\pool}{NTP Pool\xspace}
\newcommand{\vsix}{IPv6\xspace}
\newcommand{\vfour}{IPv4\xspace}
\newcommand{\eui}{\ac{EUI-64}\xspace}

\begin{document}
%\title{It's All Relative: Exhaustive and Longitudinal Characterization of the NTP Pool}
\title{On Borrowed Time: Measurement-Informed Understanding of the NTP
       Pool's Robustness to Monopoly Attacks
%       {\\ \large Not for Attribution or Redistribution}
}

\author{\IEEEauthorblockN{Robert Beverly}
	\IEEEauthorblockA{San Diego State University\\
		rbeverly@sdsu.edu}
	\and
	\IEEEauthorblockN{Erik Rye}
	\IEEEauthorblockA{Johns Hopkins University\\
		rye@jhu.edu}}
%\author{}

\IEEEoverridecommandlockouts
\makeatletter\def\@IEEEpubidpullup{6.5\baselineskip}\makeatother
\IEEEpubid{\parbox{\columnwidth}{
		Network and Distributed System Security (NDSS) Symposium 2026\\
		23 - 27 February 2026 , San Diego, CA, USA\\
		ISBN 979-8-9919276-8-0\\  
		https://dx.doi.org/10.14722/ndss.2026.240541\\
		www.ndss-symposium.org
}
\hspace{\columnsep}\makebox[\columnwidth]{}}

\maketitle

% "NTP Market power attack."
% "Monopolisation"
% "Sherman" attack

\begin{abstract}

Internet services and applications depend critically on the
availability and accuracy of network time.  The \ac{NTP} is one of the
oldest core network protocols and remains the de facto mechanism
for clock synchronization across the Internet today.  While multiple
NTP infrastructures exist, one, the ``\pool,'' presents an attractive attack
target for two basic reasons, it is: 1) administratively distributed
and based on volunteer servers; and 2) heavily utilized, including by
IoT and infrastructure devices worldwide.  We 
%develop measurements to
gather the first direct, non-inferential, and comprehensive data on
the \pool, including: longitudinal server and account membership,
server configurations, time quality, aliases, and global query traffic
load. 
%This data affords visibility into not only volunteer
%servers actively participating in the pool, but also those servers
%unavailable or evicted for providing bad time.  
%RB: too strong; i.e., the paper doesn't later discuss how we
%    find servers providing bad time.

We gather complete and granular data over a nine month period to
discover over 15k servers (both active and inactive) and shed new light into the \pool's
use, dynamics, and robustness.  By analyzing 
address aliases, accounts, and network connectivity, we find that only 19.7\% of the
pool's active servers are fully independent.  Finally, we show that an
adversary informed with our data can better and more precisely mount ``monopoly attacks'' to
capture the preponderance of NTP pool traffic in 90\% of all countries with
only 10 or fewer malicious NTP servers. Our results suggest 
multiple avenues by which the robustness of the pool can be improved.

%The \ac{NTP} is the defacto mechanism for time synchronization on the
%Internet, and significant prior literature has studied
%vulnerabilities in its design and implementation.  In contast, this
%work focuses on characterizing and analyzing a  
\end{abstract}


% \fi
\IEEEpeerreviewmaketitle

\section{Introduction}

Accurate time is critical to the function and security of distributed
systems.  The \acf{NTP} is the long-standardized and well-adopted
protocol for synchronizing time between systems on the
Internet~\cite{rfc5905}.  The security of the protocol itself has been
well-studied, with prior work demonstrating \eg time-shifting and
denial of service attacks~\cite{malhotra2015attacking, czyz2014taming,
perry2021devil}, while recent efforts standardize \ac{NTP} security
mechanisms for authentication and integrity~\cite{rfc8915}.  However,
these \ac{NTP} security mechanisms do not fully protect against malicious
time servers or availability attacks.  

In part due to outstanding security concerns, and to mitigate potential risk,
several commercial operating system vendors and network providers operate their
own NTP infrastructure, including Microsoft, Apple, Cloudflare, and
Google~\cite{windowstime,googletime}.  However, open source operating
system distributions based on Linux and BSD utilize the \pool
project~\cite{ntppool} for time synchronization by default.  In addition to
the prevalence of these operating systems in server and network infrastructure, 
embedded Linux is widely deployed on Internet of
Things (IoT) devices such as printers, webcams, WiFi routers, and home
automation.  Thus, the \pool (herein referred to simply as the
``pool'') is well-used and critical to the
operation of a large number of in-the-wild Internet devices and
services.

This work develops new measurements to shine new light on the \pool and
bring new insights into its overall robustness.
Whereas prior efforts to characterize the \pool rely on
\emph{indirect} inferences, \eg via large-scale querying of the
DNS~\cite{moura2024deep,kwon2023did,rytilahti2018masters}, we develop
a means for \emph{direct} (non-inferential) measurement.  Our method permits exhaustive
and longitudinal measurement of the pool, and affords insight into
previously unavailable information including: 1) all servers that are
members of the pool, including poor-quality, offline, and ``monitor-only''
servers not returned 
in query responses; 2) query
traffic volume and distribution; 3) server speed and configuration; 4)
aliased servers that distort the perceived diversity; 5)
accounts controlling a large number of servers; and 6) server
lifetime.  
%In addition to characterizing these data and pool
%properties, we examine their implications to the overall robustness
%and resilience of the pool.
In sum, we make the following primary contributions:
\begin{enumerate}
 \item Development and validation of a custom scraper to exhaustively and
longitudinally gather granular data on the pool, including
servers, accounts, zones, addresses, scores, traffic, and popularity
(\S\ref{sec:method}).
 \item Fingerprinting to identify and characterize NTP server
``aliases''
present in the pool, including across IP protocol versions
(\S\ref{sec:data:fp}).
 \item Measurement-based characterization of the pool,
including showing an inferred global pool rate of about 100k DNS queries
per second
(\S\ref{sec:data}).
 \item Analysis and evaluation of pool server independence, 
including account owners, network diversity, aliases,
and lifetimes,
showing
that only approximately 20\% of the active servers are independent
(\S\ref{sec:results}).
 \item Demonstration and validation of how the netspeed data we gather can be
utilized to better and more precisely mount targeted  ``monopoly attacks'' 
\cite{perry2021devil}
whereby 
the attacker captures
the preponderance
of NTP pool traffic in 90\% of all countries with 10 or fewer
attack servers
(\S\ref{sec:attack}).
\end{enumerate}

The remainder of the paper is organized as follows.  We briefly review
NTP and the \pool in \S\ref{sec:background}, while
\S\ref{sec:method} details our methodology.  \S\ref{sec:data} 
describes our datasets and important macro characteristics of the
pool.  We then investigate the degree to which participating pool
servers are independent from one another in \S\ref{sec:results}.
Next, we explore the feasibility of the monopoly attack
using our data in \S\ref{sec:attack}.  Finally, we conclude with a discussion on
ethical considerations and recommendations for improving the
robustness of the pool.

\section{Background}
\label{sec:background}

% Paragraph on general NTP stuffs; point reader to something for a
% more complete treatment
\ac{NTP}, standardized in 1985, is a protocol to synchronize system clocks among a 
distributed set of servers across a variable latency, best effort packet switched
network~\cite{rfc958,rfc5905}. NTP
organizes distributed devices into a hierarchy, rooted in a reference clock 
(stratum 0, with high-precision time sources).  Stratum 1 servers
synchronize with stratum 0 time sources, while stratum 2 servers synchronize with 
stratum 1 servers, and so on. The history and evolution of NTP over its
four-decade lifespan has been described in detail by prior
research~\cite{mills2003brief,novick2015practical,durairajan2015time,mills2017computer}.

% Paragraph on general NTP attacks; say that we focus instead on
% resilience of the pool rather than the protocol.
Significant prior work has demonstrated attacks against the NTP, \eg
time
shifting~\cite{malhotra2015attacking,annessi2017s,rytilahti2018masters,deutsch2018preventing}. 
A wide-ranging host of applications and services rely accurate time, from TLS
certificate validation~\cite{zhang2014analysis,malhotra2015attacking} to
authentication~\cite{malhotra2015attacking,kohl1993kerberos}.  It is
well-accepted that inaccurate or 
incorrect time synchronization can enable multiple attacks including
\eg denial of service or
incorrect trust calculations. 

NTP can also be used as a means for network reflection and amplification
attacks~\cite{rossow2014amplification,czyz2014taming,rudman2015characterization,jonker2017millions,kopp2021ddos,respoof-ccs19}
due to its use of connectionless UDP as its transport-layer protocol. Certain NTP message types
(such as the \texttt{monlist} request, which induces servers to provide NTP
client statistics) have amplification factors ranging from 100s to 1000s of
times the size of the request. While \texttt{monlist} in particular has been deprecated
since 2014~\cite{alert2020ntp}, other NTP queries (such as \texttt{version} and
\texttt{showpeers}) are also well-suited for amplification
attacks~\cite{czyz2014taming}.

Rather than developing new attacks, this effort instead focuses on
characterizing the \pool through new measurements and understanding the
robustness, resilience, and independence of its volunteer-operated, distributed
network of NTP servers.

\subsection{The \pool}
% Paragraph describing the NTP pool and its operation
The \pool is a system to coordinate queries to a distributed set
of volunteer-run NTP servers; in particular, the \pool itself does not operate NTP
servers.  Instead, the \pool provides a public website with
information, statistics, and the ability for volunteers to register their own
servers.  Servers are assigned to one or more ``zones'' based on their
geographic location; these zones consist of country codes and
continents, as well as a global ``@'' zone.  The availability and quality of time
provided by participating servers is monitored by the pool, via a
distributed set of dedicated ``monitors,'' to form a
``score.''  The monitoring and scoring algorithm has been described in detail in
prior work~\cite{kwon2023did,song2022analysis}. In short, however, scores are
initially set to 0 and can range from -100 to 20.  A server's score increases
when it responds with accurate time to a monitor's queries, and decreases when
it is unresponsive or provides inaccurate time.  
%Only servers with a
%score greater than or equal to 10 are included in the set of active
%servers that the pool will return to clients.

The \pool then apportions servers to clients using the Domain Name
System (DNS)
based on multiple factors, including the server's score, registered
``netspeed,'' and geographic zone.  The \pool further uses the
client's geolocation to prefer servers from the same or nearby zones.
Clients from countries with no NTP Pool servers receive DNS
responses with servers from their continent zone.
To ensure high-quality, reliable servers are provided to clients, only
``active'' servers -- those with a score greater than 10 -- are included in 
responses provided by the \pool. Server operators can influence the frequency with
which their server (and, hence, the traffic load) is provided by adjusting their netspeed value.
By using hierarchical geographic server zones, a continuous monitoring and
scoring system, and server administrator query load tuning, the \pool
is designed for
dynamic allocation, load-balancing, robustness, and resilience to failures.

\begin{figure*}[ht!]
 \centering
 \resizebox{1.5\columnwidth}{!}{\includegraphics{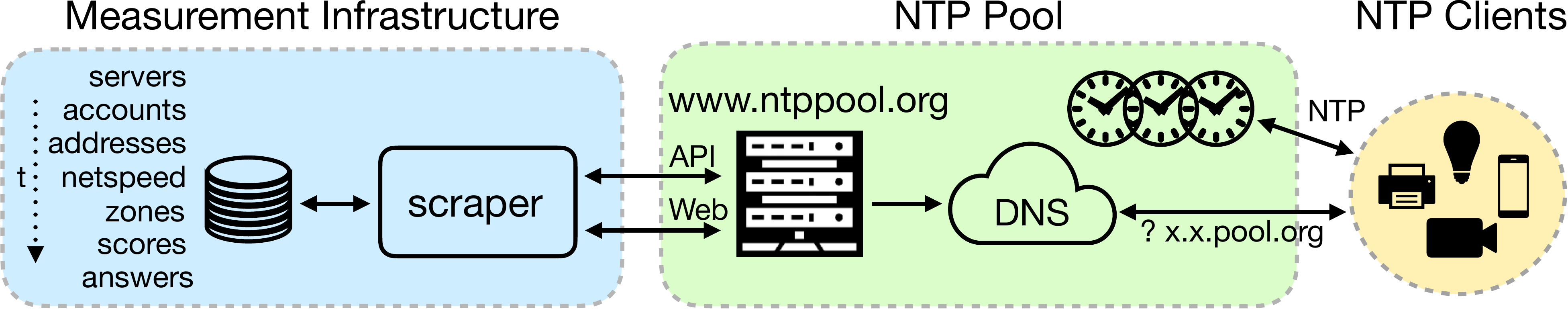}}
 %\vspace{-2mm}
 \caption{Methodology: The \pool website maintains statistics and
APIs (green box) that we periodically query (blue box) to exhaustively 
enumerate participating servers and their properties.  We gather
multiple longitudinal datasets described in Table~\ref{tab:datasets}.}
 \label{fig:scraper}
\end{figure*}

\subsection{Prior Work on The \pool}

% Paragraph on prior NTP pool measurement work
Prior work has sought to characterize the \pool.
Rytilahti \etal~\cite{rytilahti2018masters} scanned the entire IPv4
address space for responsive NTP hosts, and further used a crawler to
repeatedly query the \pool's DNS and query the returned NTP servers.
Similarly, the work of Moura~\etal subsequently used the RIPE Atlas
infrastructure to issue DNS queries against the \pool from a wider
geographic region, again to discover \pool
servers~\cite{moura2024deep}.  These studies used large volumes of 
active DNS
measurements to indirectly infer properties of the \pool.  Thus, if a
participating server is in the pool, but not served in the DNS (\eg
because it is offline, not providing incorrect time, or configured to receive no
NTP query load), these DNS-based methods will
not discover the server.  In contrast, our approach does not use the DNS,
but rather directly queries the \pool web site to produce more accurate and
complete information, as well as a rich set of additional data
including accounts,
scores, and DNS answer rates crucial to our analysis.

Most closely related to our exploration of the feasibility of
monopoly attacks against the \pool, Perry~\etal demonstrated the
potential for malicious time servers to join the \pool and 
carry out time shifting attacks by using large netspeed
values~\cite{perry2021devil}.  Whereas their approach empirically
determined the number of NTP servers an attacker would need to
impact  five large pool zones, our data affords new insights
into the broader feasibility of these attacks and the \pool's 
robustness.  In particular, we mathematically deduce the
number of servers required to monopolize the traffic of \emph{any} zone --
without needing to add servers to the zone \emph{a priori}.  

Finally, studies have used the NTP Pool as a vehicle to measure other network
properties. For instance, Durairajan~\etal used data collected by NTP Pool
server operators to measure one-way delays at scale on the
Internet~\cite{durairajan2018timeweaver}. Rye and Levin subsequently used the NTP Pool to
collect active \vsix addresses, particularly from 
clients, which are difficult to discover using active measurements~\cite{rye2023ipv6}.
And Syamkumar~\etal used data collected by NTP Pool server operators to detect
network events, such as route changes or outages~\cite{syamkumar2018wrinkles}.

\subsection{Motivation}

A key motivation of our work is the widespread use of, and dependence
on, the \pool,
effectively rendering it \emph{critical infrastructure}.  While many
desktop and mobile operating systems utilize different closed NTP
servers, \eg NTP servers operated by Apple, Google, and Microsoft, a
large number of Internet of Things (IoT) and infrastructure devices
utilize the \pool.  We base this assertion on three observations:

\begin{itemize}
  \item Data from the IPv6 Observatory~\cite{rye2023ipv6}, which collects \vsix
addresses from \pool clients, supports the NTP Pool's use by many embedded Linux
and IoT devices. For example, during the week beginning on April 20, 2025, the IPv6
Observatory was visited by over 5.5M unique Fritz!Box routers, 1.6M Amazon devices, and
289k Sonos speakers, all identified through their use of \eui
addresses in IPv6, which
embed an interface's \ac{MAC} address.  Of note, end users typically
do not (or cannot) reconfigure the chosen NTP server for such devices.
  \item Prior work from Moura~\etal~\cite{moura2024deep} examined DNS 
queries at root name servers to estimate the popularity of the \pool,
and find that, in their dataset, the \pool receives 90M of the 126M total
NTP DNS queries.  Thus, the \pool is the most popular time provider by
a large margin.
  \item As we will show in \S\ref{sec:results}, the DNS ``answer''
statistics maintained by the \pool show a global DNS query rate to the
pool of over 100k queries per second.  Given DNS caching, the number
of unique clients utilizing the \pool is orders of magnitude higher.
\end{itemize}

This large-scale use, combined with the unique volunteer nature 
of the \pool implies that traffic capture and time manipulation
attacks would be highly impactful.  Our work represents a
comprehensive characterization of the pool's robustness to such
attacks.

%I wonder if it's worthwhile to measure RTT latency for the HE connected servers (there are a ton of tunnel addresses)
%i.e, how many of these are physically elsewhere, but just using the tunnel to appear to be someone else geographically

\begin{table*}[t!]
\caption{HTTP and API endpoints provided by \texttt{https://www.ntppool.org}. 
The server ID is an internal monotonically increasing integer.  By 
querying the space of server IDs, we obtain all server IPs.}
\label{tab:endpoints}
\centering
%\resizebox{1.0\columnwidth}{!}{
\begin{tabular}{llll}
Endpoint & Parameters & Response Type & Returns \\\hline
\texttt{/scores/\{\%d\}} & Server ID & HTTP 301 & Redirect to
\texttt{/scores/\{ip\}} \\

\texttt{/scores/\{\%s\}} & Server IP & HTML & Server Info \\

\texttt{/scores/\{\%s\}/json} & Server IP & JSON & Server Scores  \\

\texttt{/api/data/server/dns/answers/\{\%s\}} & Server IP & JSON & Per-zone
DNS answers that include the server IP\\

\texttt{/api/data/zone/counts/\{\%s\}} & Zone & JSON & Per-zone
server counts and aggregate ``netspeed'' \\
\end{tabular}
%}
\end{table*}

\section{Methodology}
\label{sec:method}

Figure~\ref{fig:scraper} provides an overview of our measurement
infrastructure in relation to the \pool, while Table~\ref{tab:datasets}
summarizes the data we collect with our infrastructure.
Our methodology is primarily based on: 1) probing APIs of the public
\pool website; 2) a custom \pool website scraper; 3) an NTP server
fingerprinter; and 4) continuous measurements.  We discuss each of
these in turn, but first present the threat model.

\subsection{Threat Model}

Within the confines of the existing \pool, we consider an attacker
that seeks to capture a preponderance of NTP traffic in a country or
region.  The ultimate goal of the attack could be either passive
surveillance and monitoring (\eg collecting live IP
addresses~\cite{rye2023ipv6}), active back-scanning (\eg port scanning
live hosts)~\cite{shodanseclists,shodanscanning}, or time skew (\eg to
incorrectly influence clients' notion of time~\cite{kwon2023did}).  

We assume the adversary is capable of: 1) setting up hosts running NTP
servers physically in a particular country or region, \eg using cloud
providers, tunnels, or virtual private servers; 2) configuring both
IPv4 and IPv6 server reachability; and 3) joining the \pool and
accessing its public services, \eg by creating accounts, adding
servers, and viewing statistics.  However, we assume that the
adversary cannot circumvent the \pool's access controls, DNS and load
balancing mechanisms, or scoring algorithms.  Further, the adversary
cannot control the servers or behavior of other \pool participants.  

% https://console.cloud.google.com/bigquery?project=ntppool
% (see 'analysis/bigtable.txt' for details)
% yearly database snapshots from 2008-present
% columns:
%   server_id, monitor_id, ts, score, step, offset, rtt, leap, error
%
% SELECT SUM(row_count) FROM ntppool.ntppool.__TABLES__
%  12033113701
%   12,033,113,701
% SELECT SUM(size_bytes) FROM ntppool.ntppool.__TABLES__
%  708,200,559,363
\subsection{Historic Score Data}
\label{sec:data:bq}

The only dataset we analyze that was not collected using our 
measurement infrastructure is the historic per-server score data.
The \pool maintainers archive complete historic score data within
Google's cloud-based BigQuery, with per-year tables from 2008 to the
present day~\cite{poolbigquery,poolmonitor}.  These tables contain approximately
12B rows ($\sim$710GB) and include per-server timestamp and score rows
with a distinct server ID column. Notably, however, this data does \emph{not}
contain the IP addresses of the servers or any other meta-data.
Because of its longitudinal coverage, we use this historic data to infer
both participating server lifetimes as well as server availability 
(fraction of time the server is a member of the pool and has a score
that allows it to participate in serving queries to the pool).

%\subsection{\texttt{www.ntppool.org}}
\subsection{Scraper}
\label{sec:data:scraper}

Prior efforts to characterize and understand the \pool have relied on
issuing large numbers of DNS queries from multiple geographic
locations to discover participating servers~\cite{moura2024deep,rytilahti2018masters}.  In
addition to being inefficient, potentially inaccurate, and inducing
undue operational load on the production system, this
DNS-based probing approach cannot discover servers in the pool that
are inactive, have a low score, or are in monitor mode, as these will never be returned in a DNS response.

In contrast, we discover the ability to exhaustively enumerate all
servers, past and present, active and inactive, by directly HTTP querying
the \pool website.  For each server, the \pool website provides a
``score'' history page that plots the accuracy of the timing
information from that server as observed by a network of sentinel
monitors.  The public URL to issue an HTTP GET for these server statistic pages requires
the NTP server's IP address -- which we do not know a priori.  However, in
examining the \pool backend infrastructure, we observe that each server
is assigned a monotonically increasing integer identifier. We
then find a URL endpoint that maps (via an HTTP-level redirection) 
\pool internal server integer identifiers to
their corresponding IP address. Table~\ref{tab:endpoints} provides the specific
endpoints our scraper and measurement infrastructure query.
%RB: to the best of knowledge, these endpoints are not publicly
%    documented?
For example, to map the server with identifier 
\texttt{59105}, we HTTP query:
\texttt{ntppool.org/scores/59105} 
which returns an HTTP 301 response with the URL:
\texttt{/scores/2001:470:1f07:c21:1::123}.  Once we have this ID to
IP mapping, we can query the other endpoints to obtain score and
answer data.

% Data:
%  - (scraper.py; runs constantly): 
%     Endpoint: https://www.ntppool.org/scores/{%d}        % id
%     Redirects to: https://www.ntppool.org/scores/{%s}    % ip
%       - IP<->ID
%       - Zones
%       - Accounts (w/ caveat)
%  - (get_scores.py; runs daily):
%     For all known servers
%     Endpoint: https://www.ntppool.org/scores/{%d}/json 
%       - timestamp, step, score, monitor_id
%  - (get_answers.py; runs daily):
%     For all known servers
%     Endpoint: https://www.ntppool.org/api/data/server/dns/answers/{%s}
%       - country code, count, points, netspeed 

We leverage the relatively small and monotonically increasing integer NTP server identifier space
to create an \pool website scraper (\Cref{fig:scraper}) to enumerate all \pool servers and
then query for the next unused identifier every 90 minutes on average.  Thus, we
discover new servers shortly after they are added to the system.  In
addition, we use a separate scraper to query the website for
statistics, scores, and metadata of all servers every day.
The retrieved metadata includes the user account associated with the server, the
country and region zones served, the server's score, and the server's
netspeed.  We are also able to detect when servers are deleted.  Note
that a server may have a low score, and thus not be included in \pool
responses, but servers are only deleted if a user requests deletion,
or the server is offline or otherwise unresponsive to NTP requests for an extended period of time. 

We began collecting the \texttt{pool-scrape} data in October 2024 and continued collecting
through July 2025, representing approximately 9 months.

% nmap ntp-info (https://nmap.org/nsedoc/scripts/ntp-info.html):
%   two requests: a time request and a "read variables" (opcode 2) control message. 
% sudo nmap -oX output.xml -oN output.txt -6 -iL targets6.rand.txt -sU -p 123 --script ntp-info

\begin{table*}[!t]
\caption{Overview of datasets, sources, and tools: we gather and use the
first five datasets in
this work; the last two datasets (shaded) represent prior work and are
included for comparison.}
\label{tab:datasets}
\centering
\resizebox{\textwidth}{!}{%
\begin{tabular}{|l|c|c|c|l|}\hline
\rowcolor{black!30}\textbf{Dataset}    & \textbf{Period} & \textbf{Source} &
\textbf{Addresses (v4/v6) } & \textbf{Description} \\\hline

bq-scores (\S\ref{sec:data:bq}) & 05-Sep-2008 -- 01-Jun-2025 & BigQuery 
& 39,756 & Server scores \\\hline

pool-scrape (\S\ref{sec:data:scraper}) & 22-Oct-2024 -- 10-Jul-2025 & scraper
& 9,955 / 5,725 & Servers, accounts, and metadata scraped from public website \\\hline

pool-answers (\S\ref{sec:data:answers}) & 28-Jul-2025 & scraper 
& 3,867 / 2,228 & Per-server, per-zone DNS response rates \\\hline

server-fp (\S\ref{sec:data:fp}) & 23-Jul-2025 & fingerprinter 
& 3,967 / 2,275 & IPv4 and IPv6 aliased servers \\\hline

ntp-residual (\S\ref{sec:data:residual}) & 08-Dec-2024 -- 24-Jul-2025 & NTP
    Pool Server 
&  198,787,734 / 109,930,726  & IPv4 and IPv6 client IPs \\\hline\hline

\rowcolor{black!30}\multicolumn{5}{|l|}{\textbf{Prior Work:}} \\\hline

\rowcolor{black!10}pool-web~\cite{ntppool}         & 10-Jul-2025 & Web
page
& 3,434 / 1,926 & Aggregate counts published on public web page \\\hline

\rowcolor{black!10}deep-dive~\cite{moura2024deep} & 26-Aug-2021  --
31-Aug-2021 & Moura~\etal
& 3,056 / 1,479 & Discovered via RIPE Atlas active DNS probing \\\hline 

\end{tabular}
}
\end{table*}

\subsection{Netspeed}
\label{sec:data:netspeed}

The \pool management interface allows volunteers to specify a
``netspeed'' for each server with discrete values in the set: 
0, 512kbps, 1.5Mbps, 3Mbps, 6Mbps, 12Mbps, 25Mbps, 50Mbps, 100Mbps,
250Mbps, 500Mbps, 1Gbps, 1.5Gbps, 2Gbps, and 3Gbps.  The intent
of the netspeed setting is to allow server operators to participate
in the pool while providing a coarse-grained method to control the received query
rate.  A common
point of confusion on the discussion boards surrounds how these
netspeed settings affect the actual data rate of received NTP queries.  

Despite the data rate (\eg Mbps) labels for netspeeds,
the setting is instead a \emph{relative} weighting\footnote{The NTP Pool server
management page states that ``this speed does not mean the wire speed of your server,
it's just a relative value to other servers.''~\cite{ntppoolmanage}}. The \pool
determines the \emph{aggregate} netspeed of all servers actively
participating in the zone and then apportions a relative fraction
based on each server's netspeed.  As a result, while the netspeed
will change the received query rate, it may bear no relationship
to the true rate. 

Consider, for example, a zone with five total servers: four servers
set to a netspeed of 25Mbps and one server with a netspeed of
100Mbps.  The aggregate netspeed in this hypothetical zone is 
200Mbps.  Thus, the first four servers will each receive approximately
one eighth (12.5\%) of the total query traffic for the zone while the  
fifth will receive half (50\%).  Of course, this is an approximation
as the \pool can only control how frequently it includes a particular
server in a DNS query for a given zone, but the final traffic rate is
proportional.

\subsection{Pool DNS answers}
\label{sec:data:answers}

We find an additional \pool web server API endpoint, ``answers,''
that takes a server's IP address as input and returns a JSON object
containing a count of per-zone DNS responses, \ie how many times the
server was included in response to a client's DNS query to the pool for a given
zone.  While the JSON does not contain a timestamp, by querying the
API endpoint for 100 different servers every minute, we experimentally
determine that the \pool web server updates the returned JSON data every 30 minutes. 

Therefore, in addition to periodically probing the \pool web server
for servers and accounts, we query this answers API endpoint for all
of the active servers (those with score $\geq10$) every 30 minutes over
the course of one day on July 28, 2025 to obtain the
\texttt{pool-answers} dataset.  

\subsection{Server Fingerprinting}
\label{sec:data:fp}

A host running an NTP server application may have multiple physical or
virtual network interfaces.  These interfaces can be numbered with one
or more IPv4 and IPv6 addresses.  Whereas a single NTP application may
listen and respond to NTP queries sent to different addresses, the
pool has a strict one-to-one mapping between an address and a server,
\ie a ``server'' is an instance of an NTP server daemon bound to a 
single IP address.
We term two IP addresses with the same NTP server application as ``NTP aliases.''

The NTP protocol defines a mode for control messages~\cite{rfc5905}.
These control messages permit management and diagnosis, for instance
``read variables''~\cite{rfc9327}.  However, for privacy and security
reasons, this functionality may be disabled or blocked by the
network, especially for remote connections.  To understand the ability
to leverage NTP control messages for fingerprinting, we queried 
IPv6 addresses active in the pool in May, 2025.  Of the 
1,658
IPv6 NTP servers in the pool that respond to an NTP time client query,
only 28 (1.7\%) also respond to a read variables request.  Manual 
investigation of the responses indicates that these few responding 
servers are running quite old versions of NTP server implementations.

We therefore do not use control messages to find aliases, but rather 
implement active server fingerprinting starting with
the open-source ``ntpdedup'' code~\cite{ntpdedup}. (We forked this
codebase which we keep anonymous for submission, but will make 
public and will contribute our modifications back via merge
requests.)

% We therefore utilize unique features of standard NTP time (mode 4)
% response packets to identify aliases.  In particular, from all
% responsive servers, we gather the NTP: version, stratum, poll interval,
% precision, reference ID, and reference timestamp.  In our 
% experimental testing, we discover that different physical stratum 1
% servers may have the same reference clock (\eg GPS) with identical
% reference timestamps; this can lead to NTP alias false positives. 
% For this reason, we do not attempt to de-alias stratum 1 servers.

Instead, we leverage unique features and fields of standard NTP time (mode 4)
response packets to provide a fingerprint and identify aliases. In particular, our modified version of 
the fingerprinting tool ntpdedup collects several NTP time response fields that
identify the server's version, time source,
the server's last synchronization time, and its precision and
maximum allowable polling interval.  The time source data includes
the reference identifier (``refid''), stratum,
and dispersion.  The refid is a 32-bit field that identifies the
reference clock for stratum-1 servers (analyzed in
\S\ref{sec:data:clocksources}) or the
synchronization peer IP address for stratum-2 and higher servers.  The
stratum field is a single byte; we identify 7 unique stratum values
among NTP servers in our data.  Similarly, polling interval and
precision fields are each one byte.  While there are only 11 unique
poll and 22 unique precision values within our data, these fields are 
static and provide course-grained differentiation to identify clear
non-aliases. 

The ``reference timestamp'' contains the time since the system's clock
was last set or corrected and is represented in ``NTP timestamp
format:'' seconds since January 1, 1900 with 32 bits encoding the
integer component of seconds and 32 additional bits encoding the
fractional seconds.  While NTP server applications synchronize time,
their individual internal update intervals are not -- hence the
precision of this field affords strong discrimination power and,
naturally, is the field that exhibits the largest number of unique
values in our data.  

When these fields are collected from a
server over a short time interval, they will be consistent, even across
different aliases of the same physical server. Thus, we can identify potential
aliases among NTP Pool servers by comparing whether these values are equal
across multiple server responses.

Our alias resolution technique can suffer from both false positives and
false negatives. Two non-alias server IP addresses are more likely to be incorrectly identified as
aliases if they synchronize with the same upstream NTP server; this problem is
particularly acute for stratum 1 NTP servers, which synchronize time with a
limited number of stratum 0 time sources (\eg GPS) and
therefore have a small set of refids. We examine the distribution of
stratum 1 server refids in
\S\ref{sec:data:clocksources}.  For this reason, and to focus on
alias precision rather than recall, we do not attempt to
de-alias 830 active stratum 1 NTP servers. Conversely, two server IP
addresses may be erroneously misidentified as non-aliases. For example, in the
event that an NTP server resynchronizes its time between queries to two IP
addresses assigned to it, its reference timestamp would differ in the two NTP
responses.  We use ground-truth servers to validate our fingerprinting
method in \S\ref{sec:method:validation} and further use account
owner and ASN as proxies to evaluate the discovered aliases across the
entire pool in \S\ref{results:aliases}.
%forward pointer to validation next section
 
Other potential NTP protocol features are
available for de-aliasing, such as the root delay and root dispersion; however, these
can vary significantly between probes and cause false negatives.
Similarly, IP-layer features such as IPv4 TTL and DSCP, and  IPv6
hop limit and traffic class, can provide intra-protocol alias hints,
but are too unreliable as discriminators, especially for identifying
cross-protocol aliases. We added collection and correlation of these weak identifiers, as
well as the server's polling interval, to ntpdedup, but do not use
them by default. To obtain a weaker alias
inference for stratum 1 servers, we could omit the refid field as part of the
alias determination. However, we opted to exclude stratum 1 servers from alias
consideration rather than include false positives.

\subsection{Residual NTP Client Traffic}
\label{sec:data:residual}

The NTP Pool warns prospective operators that running an NTP server as part of
the pool is a long-term commitment. The pool further indicates that a
server operator may continue to receive NTP traffic to their servers ``weeks,
months, or even YEARS before the traffic goes completely away'' due to NTP
server IP address caching~\cite{ntpjoin}. 
This means that
an adversary that joins the NTP Pool to receive potential victims in the form of
NTP clients may be able to continue to attack those victims (\eg by shifting
their clocks) even if detected and evicted from the pool.

In order to understand the quantity and duration of client requests to a
pool server after it has been evicted, we conduct some small-scale experiments
with an NTP server under our control. We add and remove this server from
the NTP Pool, but continue to monitor received traffic levels for months after
it was evicted. This helps us understand how long and how many clients may be at
risk even in the event that a malicious NTP server is removed from the \pool.

\subsection{Validation}
\label{sec:method:validation}

For the purposes of validation, we configured eight virtual private
servers (VPS) in various geographic regions.  We added both the
IPv4 and IPv6 address of one of the VPS to the \pool.  For the
remaining seven VPSes, we configured two IPv6 addresses each and
added them to the \pool.  Thus, we added a total of 16 servers
to the \pool.  We then examined the operation of our scraper
and measurement infrastructure by validating: 1) presence of
the server in our database; 2) correct zones and meta-data; and 3) the
inferred server age, \ie the duration of time it was participating
in the \pool.  For all 16 servers, across these three metrics,
we achieve perfect accuracy, providing an additional degree of 
confidence in the correctness and completeness of the data we gather.

We next evaluated the accuracy of our NTP server fingerprinting
code in identifying NTP aliases.  In addition to the eight different pairs of
aliases in our ground truth set of servers (one mixed IPv4/IPv6
alias, and seven IPv6 pair aliases), we selected 100 active servers 
with a score $\geq$ 10 at random.  We then used our fingerprinter to probe all 
116 addresses.  Within this sample experiment, 
our alias detection correctly identified all eight aliases and 
did not produce any additional false positive aliases. %-- suggesting perfect 
%recall and precision.  

On one of our ground truth VPSes, we then added 10 different IPv6
addresses to a single interface and again ran our fingerprinting
code.  The fingerprinter correctly identified this cluster of 10
addresses as belonging to a single server.  Finally, we experimented
with running stock configurations of two popular NTP server
applications, 
\texttt{ntpd} and \texttt{chronyd}, both running on the same physical
VPS, but with \texttt{ntpd} bound and listening to two different IPv6
addresses and \texttt{chronyd} bound to a third address.  In this
instance, our fingerprinter identified one cluster of two addresses
and one cluster of a single address, indicating that NTP server implementation
plays a role in cluster determination.  
% /etc/ntpd.conf:
%  interface ignore wildcard
%  interface listen 2001:470:1f07:c21:1:0:1:12b
%  interface listen 2001:470:1f07:c21:1:0:1:12c
% /usr/local/etc/chrony.conf
%  bindaddress 2001:470:1f07:c21:1:0:1:12a
%  allow ::/0
%  local stratum 4

\subsection{Limitations}
\label{sec:method:limits}

While we believe our dataset represents the most accurate 
characterization of the \pool to-date, we note several 
limitations of our methodology.  First, the public webpages
for each server do not contain account names or organization
information in cases where the user has configured their 
account to be private.  While we discover 1,332 unique
accounts and 15,680 servers, we are only able to infer the
accounts responsible for 7,225 (46\%) of the servers.  Thus,
our account visibility is limited.

Second, while the BigQuery data include historic scores
dating back to 2008, this data can only be used for inferring
server lifetimes.  The remainder of our analyses are based on
the scraping and measurement infrastructure which includes only
9 months of longitudinal data.  

Finally, while we perform multiple experiments to validate 
the accuracy of our fingerprinting, we note that corner cases
may exist that lead to false positive or false negative aliases.
For instance, in the case that two different NTP server application 
implementations 
are running on different interfaces of the same physical machine,
our fingerprinter cannot ascertain that these are aliases.  
However, we believe such instances to be uncommon.

\begin{table}[t!]
\caption{\pool Summary Statistics (pool-scrape dataset) as of July 10, 2025}
\label{tab:poolstats}
\centering
\begin{tabular}{|l|r|r|}\hline
\rowcolor{black!30}  & \textbf{IPv4} &\textbf{IPv6}\\\hline
Servers (IP addresses)              & 9,955 & 5,725 \\\hline
Autonomous Systems   & 2,107 & 841 \\\hline
Zones                & 29 & 29 \\\hline
Active Servers       & 3,967 & 2,275 \\\hline
Servers w/ Accounts  & 4,277 & 2,948 \\\hline
Stratum 1 Servers    & 548 & 282 \\\hline
Monitor-only Servers & 1,672 & 1,007 \\\hline
Anycast Servers      & 7 & 5 \\\hline
\end{tabular}
\end{table}

\section{Pool Characterization}
\label{sec:data}

Because \pool membership requires only that a volunteer have a publicly
accessible NTP server, many types of individuals, institutions, and
organizations might consider adding their server. To characterize the types of
servers and server operators that make up the \pool, we analyze both current and
historical IP addresses that comprise the \pool as well as the accounts linked
to those IPs over several axes.

\subsection{Server Lifetimes}

\begin{figure}[t!]
        \centering
        \resizebox{0.9\columnwidth}{!}{\includegraphics{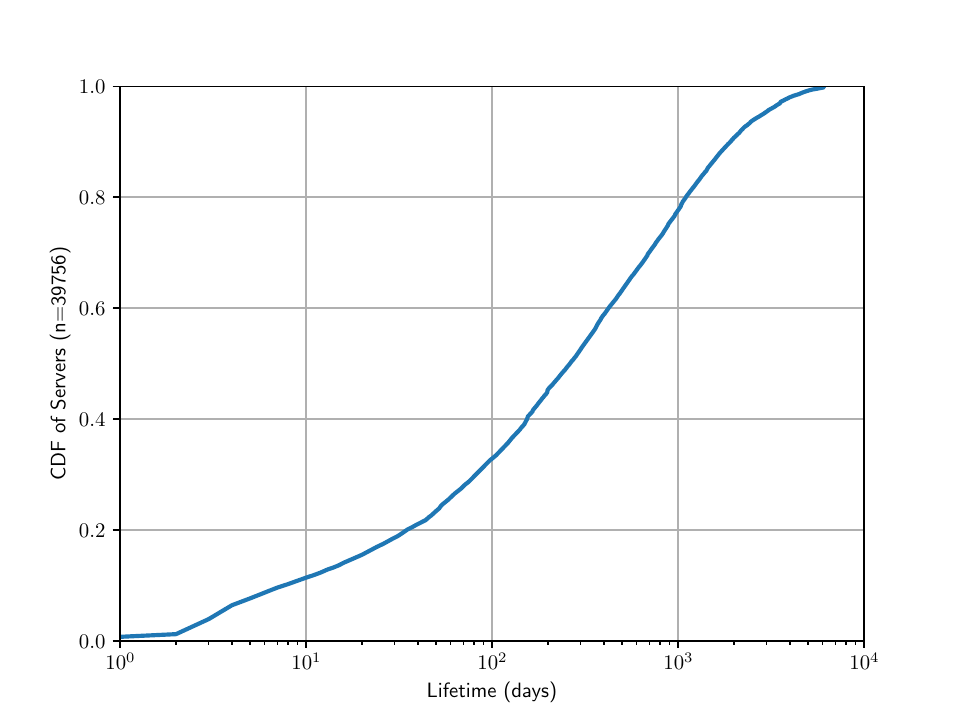}}
        \caption{Lifetime of \pool servers.
                 More than 10\% of servers participate for less than 10 days.}
        \label{fig:lifetime}
\end{figure}

We first consider the ``lifetime'' of the servers in the pool and
analyze the \texttt{bg-scores} data covering 17 years and nearly 
40,000 servers.  \Cref{fig:lifetime} shows the cumulative fraction
of servers as a function of their lifetime, as inferred by the
presence of scores in the dataset.  We see that the median 
lifetime is approximately one year, while more than 10\% of servers
participate for less than 10 days.  Conversely, approximately 20\%
of the servers have been participating in the pool for 3 or more
years.  Based on this analysis, the pool might consider prioritizing
returning servers with higher uptimes to improve the pool stability,
as well as mitigate short-lived or ephemeral attacks.

\begin{figure}[t!]
    \centering
    \resizebox{0.8\columnwidth}{!}{\includegraphics{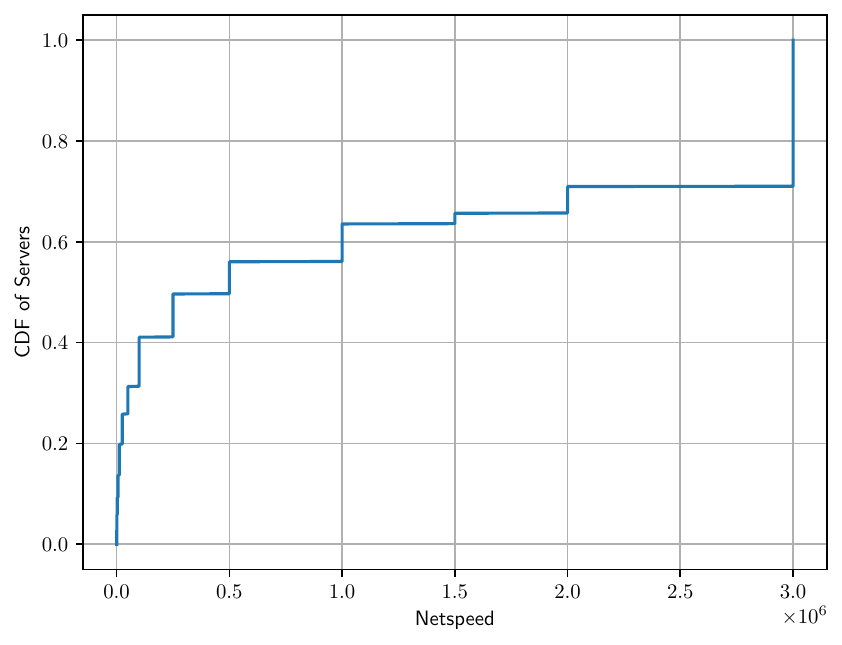}}
    \caption{Distribution of the netspeeds of the 5,333 servers with nonzero
    netspeed in the \pool. Another 57,049 servers have zero netspeeds, either
    because they are set to ``monitor-only'' or have been deleted.}
    \label{fig:netspeeds}
\end{figure}

\subsection{Clock sources}
\label{sec:data:clocksources}

As described in \S\ref{sec:background}, each NTP server synchronizes with 
servers at a lower stratum and serves servers with higher stratum.
Within our data, we find 830 stratum 1 servers, 548 of which are
IPv4 and 282 IPv6.  Stratum 1 servers depend on stratum 0
high-precision time sources for their reference clock.  Per the
NTP protocol, the reference identifier (``refid'') in NTP packets from 
stratum 1 servers encode their clock source in ASCII~\cite{ntpcodes}.  
\Cref{tab:stratum} provides the distribution of the top 10 most
common reference clock identifiers.  Global Navigation Satellite
System (GNSS) sources are by far the most common, including the
PPS, GPS, and GNSS identifiers.  However, there is a large range
of identifiers with 59 unique refids across all servers, 
many of which are not standardized.  

The small number of false positives we find in our NTP server dealiasing
(\S\ref{sec:data:fp}) are due to collisions
among stratum 1 servers caused by the limited number of refids.  We therefore exclude stratum 1 servers from
the dealiasing component of our analysis.

\begin{table}[t!]
\caption{Top 10 Most Common \pool Stratum 1 Server Reference Clocks}
\label{tab:stratum}
\centering
\begin{tabular}{|l|r|r|}\hline
\rowcolor{black!30} \textbf{Clock} & \textbf{IPv4 Count} &
\textbf{IPv6 Count}\\\hline
PPS & 200 (36.5\%) & 105 (37.2\%) \\\hline
GPS & 151 (27.6\%) & 63 (22.3\%) \\\hline
GNSS & 25 (4.6\%)  & 16 (5.7\%) \\\hline
MRS & 19 (3.5\%)   & 10 (3.5\%) \\\hline
PPS0 & 18 (3.3\%)  & 14 (5.0\%) \\\hline
MBGh & 15 (2.7\%)  & 0 (0.0\%) \\\hline
PTP0 & 9 (1.6\%)   & 14 (5.0\%) \\\hline
PHC0 & 6 (1.1\%)   & 2 (0.7\%) \\\hline
kPPS & 6 (1.1\%)   & 3 (1.1\%) \\\hline
DCF & 5 (0.9\%)    & 2 (0.7\%) \\\hline
\end{tabular}
\end{table}

\subsection{Netspeeds}

The ``counts'' API endpoint (see Table~\ref{tab:endpoints}) provides
a direct means to query the pool on a per-zone basis for the 
count of active IPv4 and IPv6 servers, as well as the \emph{aggregate}
netspeed.
Using our ground-truth servers in sparsely populated zones, we
modified the server's netspeed and experimentally verified the 
effect on the reported aggregate netspeed, as well as the received
query volume.

Figure~\ref{fig:netspeeds} displays a CDF of the 5,333 \pool
servers with nonzero netspeed. More than half of all NTP servers have a netspeed
of 500 Mbps or less, which may permit an attacker with a significantly higher
rate to dominate a zone's NTP queries (\S\ref{sec:attack}).

\subsection{Traffic Load}

Our \texttt{pool-answers} dataset allows us to compute the
aggregate rate of DNS responses the \pool returns for different zones.
Note that this rate is distinct from the number of NTP queries
arriving at a pool server or the servers within a zone due to DNS
caching effects.  Thus, the rates we compute are a strict lower bound
of the total number of actual NTP queries.

\begin{figure*}[t]
 \centering
\begin{subfigure}{0.5\textwidth}
  \centering
  \includegraphics[width=.9\linewidth]{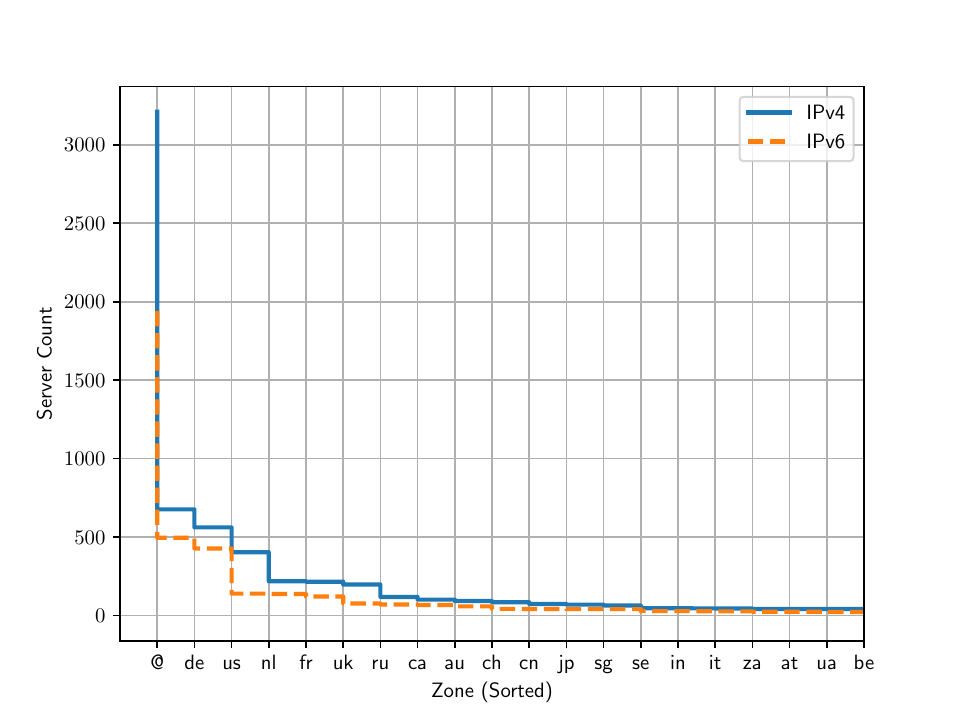}
  %\vspace{-2mm}
  \caption{Number of servers within each pool zone}
  \label{fig:dns:servers}
\end{subfigure}%
\begin{subfigure}{0.5\textwidth}
  \centering
  \includegraphics[width=.9\linewidth]{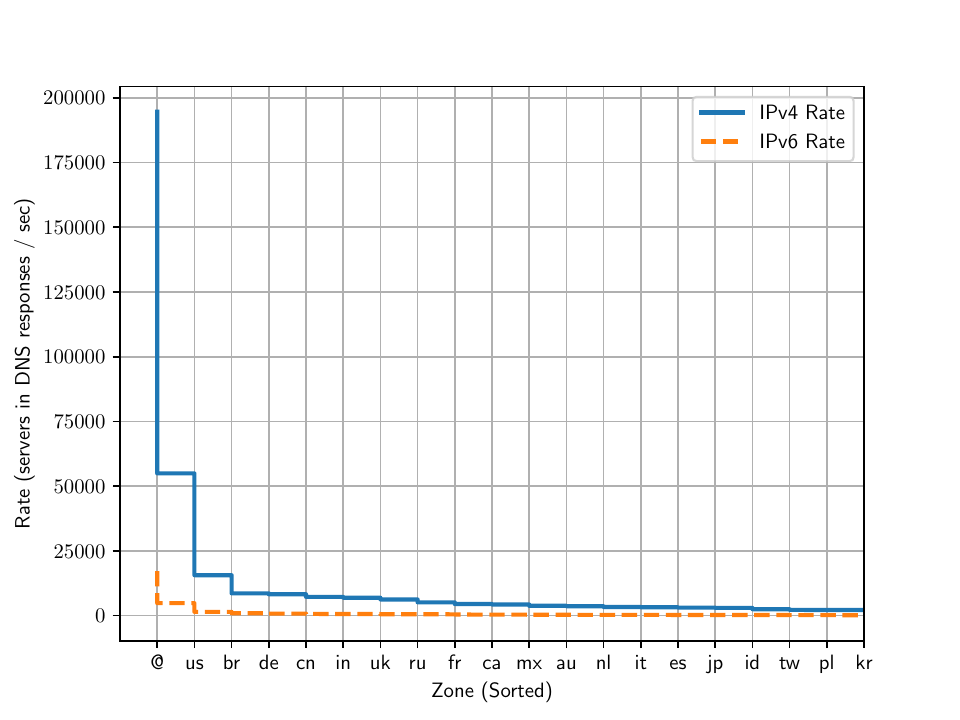}
  %\vspace{-2mm}
  \caption{Inferred DNS response rates across NTP pool zones}
  \label{fig:dns:rate}
\end{subfigure}%
 \caption{Pool DNS response statistics: both the distribution
  of servers across zones as well as the per-zone DNS answer rates
  are highly skewed.}
 \label{fig:dns}
\end{figure*}

\Cref{fig:dns:servers} displays the number of servers participating
in each pool zone for both IPv4 and IPv6.  Note that a server may
belong to more than one zone, hence the total sum of counts in the
plot is larger than the number of active servers.  Consistent with
prior work, we also find that the distribution of servers across 
zones is highly skewed and that many zones remain underserved.  As
a result, the global ``@'' zone experiences a very high query rate
and the operators have sought to include anycast servers in the pool.
We analyze anycast servers in the next subsection.

\Cref{fig:dns:rate} provides the distribution of inferred DNS 
response rates per zone, while
Table~\ref{tab:rate} shows the number of servers for the top 10
highest DNS answer rate zones, where ``@'' is the global zone.  We
estimate that the entire pool system is returning approximately 
390k and 34k IPv4 and IPv6 servers in DNS responses per second.  Since
the pool returns up to four addresses per DNS query, this equates to 
a global rate of approximately 106k queries per second.

We see an order of magnitude higher rate for IPv4 as 
compared to IPv6, likely due to the fact that the pool will only 
return a DNS response containing an IPv6 server if the
``2.xxx.pool.ntp.org'' name is queried.  We see that the global 
zone and the United States and Brazil zones account for approximately
68\% of all DNS answers.
%TODO: needs cite

\begin{table}[t]
\caption{\pool DNS answer aggregate rates across the top 10 zones. 
While the rates vary significantly by zone, we infer a large 
    aggregate system rate of over 100k DNS queries / second (each DNS response
    contains up to 4 NTP servers).}
\label{tab:rate}
\centering
\begin{tabular}{|c|c|c|c|c|}\hline
\rowcolor{black!30} & \multicolumn{2}{|c|}{\textbf{IPv4}}  & \multicolumn{2}{c|}{\textbf{IPv6}} \\\hline
\rowcolor{black!30}\textbf{Zone} & \textbf{Servers} & \textbf{Rate
(servers/sec)} & \textbf{Servers} & \textbf{Rate (servers/sec)} \\\hline
@  & 3,211 & 194,653 & 1,941 & 17,201 \\\hline
us & 677 & 54,924 & 428 & 4,804 \\\hline
br & 29 & 15,608 & 20 & 1,454 \\\hline
de & 562 & 8,599 & 496 & 914  \\\hline
cn & 34 & 8,238 & 43 & 768 \\\hline
uk & 216 & 7,235 & 122 & 611 \\\hline
ru & 404 & 6,917 & 77 & 509 \\\hline
in & 45 & 6,244  & 29 & 658 \\\hline
fr & 219 & 5,122 & 138 & 487 \\\hline
ca & 119 & 4,473 & 71 & 402 \\\hline\hline
Total & 3,867 & 389,257 & 2,228 & 34,399 \\\hline
\end{tabular} 
\end{table}

\subsection{Anycast}

Among the servers we discover in \texttt{pool-scrape}, we find 12
registered in two or more different continent zones.  Seven of these
servers have IPv4 addresses, while five have IPv6 addresses.  Two of
these servers are located in T\"{u}rkiye and are in both the Asia and
Europe continent zones, which is consistent with T\"{u}rkiye's physical
location.   Four servers, consisting of two IPv4/IPv6 aliases belong
to the same account -- a regional ISP.  The remaining six servers are
all IP anycast, as determined by looking up their addresses in the
MAnycast anycast census~\cite{hendriks2025manycastreloadedtoolopen}.
Four of these anycast servers belong to Cloudflare, while two are
within a Dutch academic network.

%194.27.222.5       time.ume.tubitak.gov.tr
%129.250.35.250     Jared Mauch
%129.250.35.251     Jared Mauch
%162.159.200.1      Cloudflare      ANYCAST
%162.159.200.123    Cloudflare      ANYCAST
%194.0.5.123        any.time.nl     ANYCAST
%176.235.250.150    ntp1.sonet.com.tr

%2001:418:3ff::53   Jared Mauch 
%2001:418:3ff::1:53 Jared Mauch
%2606:4700:f1::123  Cloudflare      ANYCAST
%2606:4700:f1::1    Cloudflare      ANYCAST
%2001:678:8::123    any.time.nl     ANYCAST

We again use the \texttt{pool-answers} dataset to examine the rate of
NTP DNS answers for these multiple continent anycast servers.  We find
that the IPv4 anycast servers are returned at a rate of 41.8k DNS responses
per second (approximately 11\% of the global traffic), while the
five IPv6 anycast servers are returned at a rate of 4.0k DNS responses
per second (approximately 12\% of the global traffic).  In
investigating these anycast servers, we find an online discussion
including the pool operators who emphasize that these servers help
provide service for zones with few or no servers~\cite{cloudfare-discuss}.  Thus, this
small set of servers are disproportionately important in the pool.

\begin{figure}[H]
        \centering
        \resizebox{0.8\columnwidth}{!}{\includegraphics{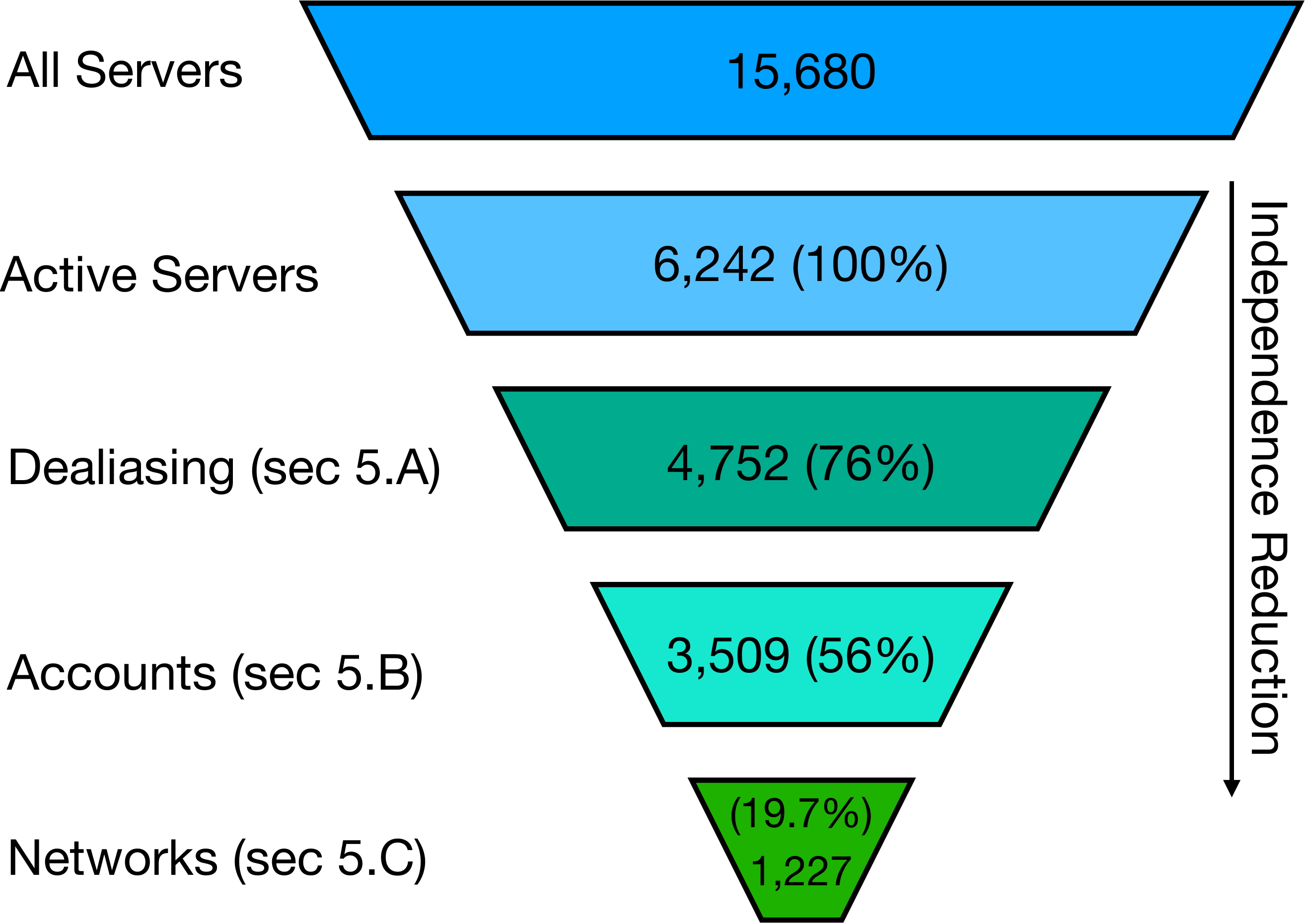}}
        \caption{Overall Independence analysis}
        \label{fig:independ}
\end{figure}

\section{Server Independence}
\label{sec:results}

In this section, we analyze the structure and composition of
the \pool as inferred from our measurements.  The primary contribution of this
analysis is to demonstrate the ways that the seemingly large number of
servers within the pool \emph{are not independent.}  In other words, we
find that many of the servers have correlated behaviors and share fate
in the event of a failure.  To better understand correlations between
servers and how servers may share fate, we employ a series of
independence reduction steps using tools and data described in the
previous sections.  

As illustrated in \Cref{fig:independ}, we begin with the full set of
servers from the \texttt{pool-scrape} dataset.  We then consider
the snapshot of 6,242 servers that had a score $\ge10$ on July 10,
2025 and count this set as the full set (100\%) of servers available
on that day.  We then use our
fingerprinting method to dealias these to unique hosts
(\S\ref{results:aliases}).  Next, we examine account uniqueness in
detail to further winnow the set (\S\ref{results:control}).  Finally,
we examine the server network connectivity and \acp{ASN}
(\S\ref{results:networks}).  As we will show, while our dataset
includes 6,242 active pool servers, only 1,227 of these are fully
independent servers -- a reduction of more than 80\%.

%\begin{table}
%\caption{Overall Independence analysis}
%\label{tab:independ}
%\centering
%\begin{tabular}{|l|l|l|l|}\hline
%\rowcolor{black!30} & \textbf{Section} & \textbf{Count} &
%\textbf{Percentage} \\\hline
%%All known servers & \S\ref{sec:data:scraper} & 15,680 & 100.0 \\\hline
%%After dealiasing & \S\ref{results:aliases} & 14,191 & 90.5 \\\hline
%%%After removing tunnels & & 13,863 \\\hline
%%After ASN reduction & \S\ref{results:networks} & 2,194 & 14.0 \\\hline
%%After account uniqueness & \S\ref{results:control} j& 1,961 & 12.4 \\\hline
%All known servers		& \S\ref{sec:data:scraper} & 15,680 & 100.0 \\\hline
%After dealiasing		& \S\ref{results:aliases}  & 14,191 & 90.5 \\\hline
%After account uniqueness	& \S\ref{results:control}  & 9,076 & 57.9 \\\hline
%After ASN reduction		& \S\ref{results:networks} & 1,934 & 12.3 \\\hline
%\end{tabular}
%\end{table}

\subsection{Server Aliases}
\label{results:aliases}

Recall that a ``server'' registered within the pool is identified by
its IP address.  Thus, the IPv4 address and the IPv6 address of a
server may correspond to the same physical machine, or the same
physical machine may have multiple IP addresses within the same
protocol family assigned to its interfaces.  In this subsection, we
turn to identifying these ``NTP aliases'' using our fingerprinting
method of \S\ref{sec:data:fp}.

We define ``NTP aliases'' as two IP addresses that respond to NTP
queries from the same NTP daemon running on a single host.  Let an
``NTP alias cluster'' be the set of IP addresses that are NTP aliases,
and the ``cluster size'' be the number of addresses in the alias set.
Let a ``singleton cluster'' be a single IP address with no identified
aliases, \ie a cluster with size one.
Last, we refer to the cluster ``covering prefix length'' as the longest IPv4 or
IPv6 network prefix mask such that the prefix encompasses all IPv4 or
IPv6 addresses within the set.  For example, if an alias cluster 
contains three IPv4 addresses: 
\texttt{1.2.1.10}, 
\texttt{1.2.3.200}, 
\texttt{1.2.14.30}, the covering prefix is \texttt{1.2.0.0/20} and
the covering prefix length is 20.

We probed 6,242 servers (3,967 IPv4 and 2,275 IPv6) within the
\texttt{pool-scrape} data with score $\geq10$ on July 23, 2025.
Of these, a total of 5,687 responded (91.1\%) to our fingerprinter.
We identify 4,123
NTP alias clusters, 2,860 (69\%) of which are singletons (\ie a
cluster with a single address and no aliases).  We exclude from
analysis 74 clusters containing one or more stratum 1 servers (a total
of 160 addresses).  

\begin{figure}[t!]
        \centering
        \resizebox{0.9\columnwidth}{!}{\includegraphics{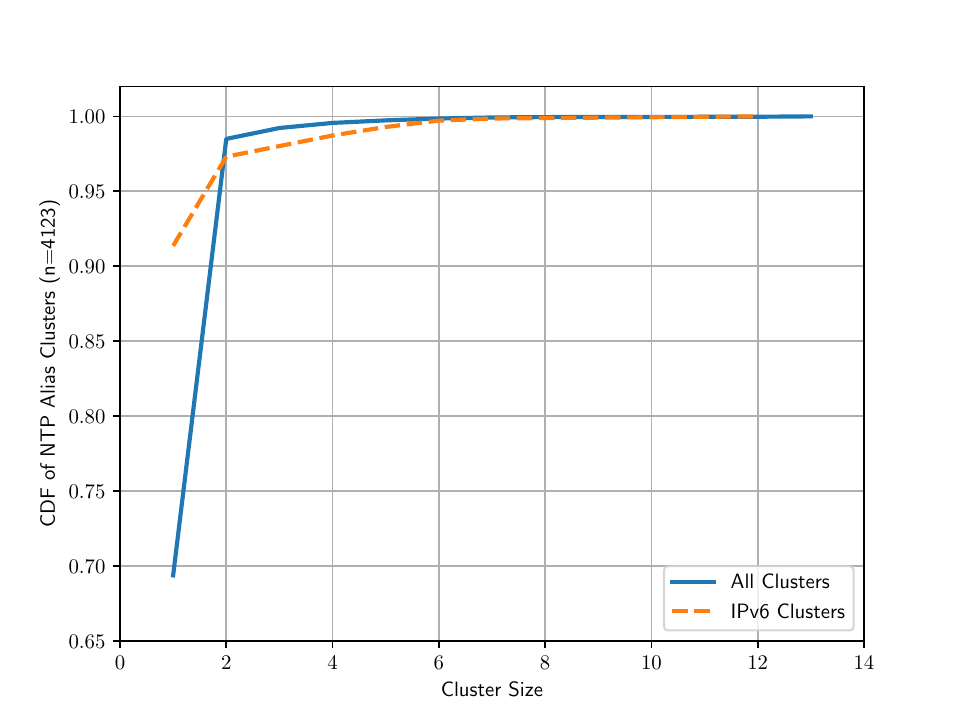}}
        \caption{NTP Alias Clusters: while 69\% of addresses belong
        to a singleton cluster, there are a significant number of
        clustered IPv4 and IPv6 pairs, and there
        exist large IPv6 clusters.}
        \label{fig:sybil_cluster}
\end{figure}

\Cref{fig:sybil_cluster} displays the
cumulative fraction of clusters as a function of their size, both
for all clusters and IPv6-only clusters.  
29\% of
the clusters are of size two, and 90\% of these consist of an IPv4
and IPv6 pair.
However, the
distribution has a long tail; for instance we find larger
IPv6 clusters containing as many as 13 addresses.

\begin{figure}[t!] 
	\centering
	\resizebox{0.9\columnwidth}{!}{\includegraphics{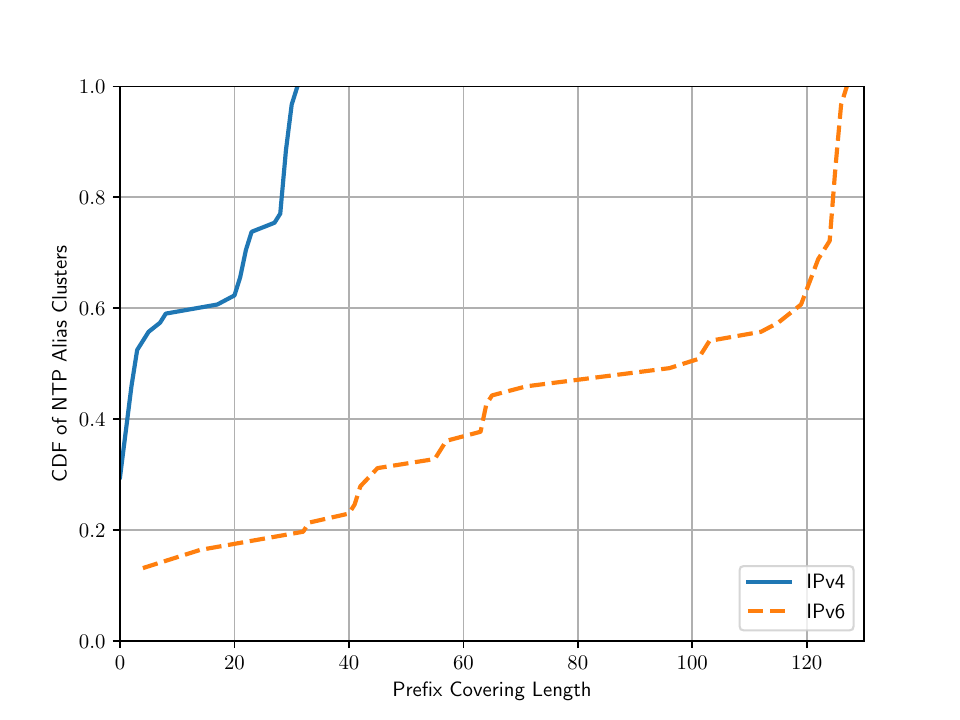}}
	\caption{NTP Cluster Covering Length (Clusters of size $\ge2$): 50\% of the addresses
        in IPv6 alias clusters
        are within the same /64.} 
	\label{fig:sybil_cover} 
\end{figure}

We then examine cluster covering sizes; \Cref{fig:sybil_cover}
displays the CDF of alias clusters as a function of their cover size
for IPv4 and IPv6 (note, it is not possible to obtain a prefix that
covers the mixed protocol clusters).  Approximately 40\% of IPv4
clusters have a covering prefix length of /20 or longer (more
specific), while the 50\% of IPv6 clusters have a covering prefix length of
/64 or longer.  
 
While this largely maps to our intuition that aliases in the same
address family should be numerically close, as they represent
addresses on the same host, the smaller prefix covering lengths
were unexpected.  To better understand these results and measure inter-cluster
consistency, we 
examined both the account owner and the \ac{ASN} to which the
servers in the clusters belong.

Among the 1,264 non-singleton clusters, we find that, for all servers within the
cluster, 1,160 have consistent, \ie matching, account owners and
\acp{ASN}.  A further 90 have consistent account owners, but
inconsistent \acp{ASN}, and 7 have matching \acp{ASN} but 
inconsistent account owners.  Only 6 clusters have both inconsistent
\acp{ASN} and account owners.  Thus, we believe our cluster inferences
to be largely correct.  Among the inconsistent \acp{ASN}, we find
that some providers have two different \acp{ASN}
for their IPv4 and IPv6 networks.  Even within the same
\acp{ASN}, that network may own and advertise multiple different
prefixes that are not numerically close in the address space.  
Finally, we find examples of IPv6 tunnels from third-party providers
that account for some of the inconsistent \acp{ASN} as well as small
covering prefix lengths.  Among the inconsistent account owners,
we discover instances of accounts with different names that are,
however, clearly related.  For instance, one account uses an
individual's full name, while a second account uses the individual's
abbreviated name.  

% Examples:
% 329: ---------------------------------
%        2001:470:6d:b7f::1 (BY) eevtsmbkj7na5 bars@hbars.by 18.9 6939 4 2 3232236387
%        2a03:e2c0:4da7:5555::1 (RU) bm27q7kbur3fsb bars@hbars.site 20.0 205125 4 2 3232236387
% 316: ---------------------------------
%        2001:858:2:4:887d:c7ff:fef8:d02d (AT) b7mnmzkcuttceibwgysaj o.polterauer@mediainvent.com 20.0 8437 4 2 1484121454
%        86.59.80.170 (AT) opb Oliver C. Polterauer 20.0 8437 4 2 1484121454
% 463: ---------------------------------
%        2001:67c:21bc:68::1 (BG) don Atanas Vladimirov 20.0 200533 4 3 1863548049
%        78.130.168.61 (BG) don Atanas Vladimirov 20.0 9070 4 3 1863548049

\subsection{Server Control}
\label{results:control}

\begin{figure}[t!]
        \centering
        \resizebox{0.9\columnwidth}{!}{\includegraphics{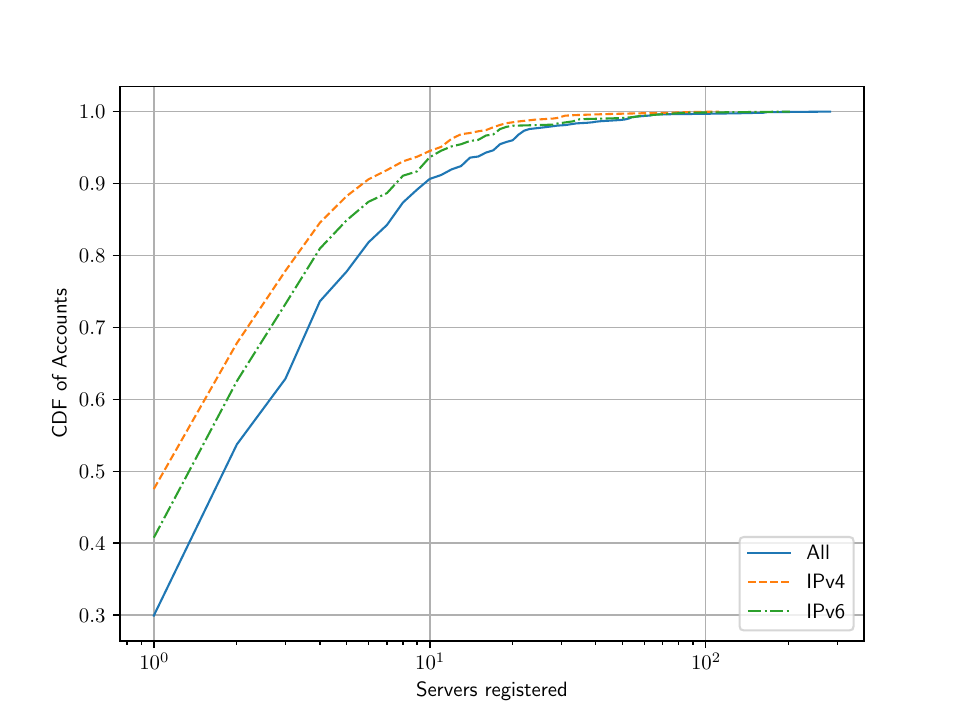}}
        \caption{Distribution of the number of \pool servers
registered per account.  A small number of accounts control a
disproportionate number of servers.}
        \label{fig:acctdistrib}
\end{figure}

While the \texttt{bg-scores} dataset provides an extended historical
view, the data only includes server IDs and scores.  We next turn to
our gathered \texttt{pool-scrape} dataset to examine accounts 
registering and controlling the participating servers.
\Cref{fig:acctdistrib} displays the cumulative distribution of
accounts as a function of the number of servers they control, broken
into IPv4 and IPv6 servers.  

%SELECT a.Name, COUNT(a.Name) AS cnt FROM Servers s, Accounts a WHERE a.AcctId=s.AcctId GROUP BY a.Name ORDER BY cnt;
While the median number of servers per
account is approximately two, the distribution has a long tail.  In
particular, we find one account that controls over 340 servers, while the
top 10 accounts control nearly 1,300 servers in total -- a significant
overall fraction of the entire pool's active servers.  Note that this
concentration is a lower bound; as noted in the limitations
(\S\ref{sec:method:limits}),
we are unable to map anonymous servers to their account owners.

%\begin{figure}[t!]
%        \centering
%        \resizebox{0.7\columnwidth}{!}{\includegraphics{acctdistrib_active}}
%        \caption{Distribution of the number of active (score>15) \pool servers
%registered per account}
%        \label{fig:acctdistrib_active}
%\end{figure}

\begin{table}[t]
\caption{\pool Server Classification: the majority of servers in the
pool are within cloud hosting or service provider infrastructures.}
\centering
\begin{tabular}{|c|r|r|}\hline
\rowcolor{black!30} \textbf{AS Type} & \multicolumn{1}{c}{\textbf{Count}} &
    \multicolumn{1}{|c|}{\textbf{\%}} \\
\hline
Hosting          & 8,500                              & 54.2                            \\
\hline
ISP              & 6,036                              & 38.5                            \\
\hline
Education        & 476                                & 3.0                             \\
\hline
Business         & 395                                & 2.5                             \\
\hline
Unknown          & 189                                & 1.2                             \\
\hline
Government       & 84                                 & 0.5				\\
\hline
\end{tabular}
\end{table}

\subsection{Server Networks}
\label{results:networks}

\begin{figure}[t!]
        \centering
        \resizebox{0.9\columnwidth}{!}{\includegraphics{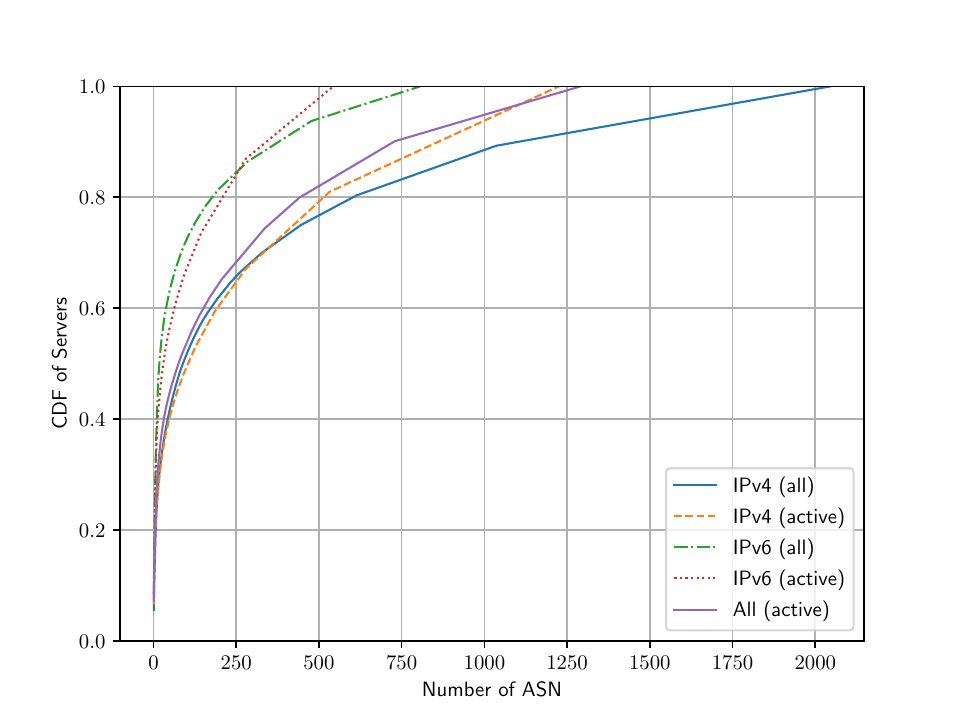}}
        \caption{Server AS distribution: the majority of servers are
         located within a small number of networks, while IPv6 does not
         substantially contribute to the network diversity.}
        \label{fig:netdistrib}
\end{figure}

We next consider the networks of the participating pool servers by
mapping the IP addresses of all of the servers we discover in
\texttt{pool-scraper} to the Autonomous System (AS) to which they
belong.  For this, we utilize a complete Routeviews BGP table
snapshot from July 23, 2025~\cite{rv} and perform longest prefix
matching.  \Cref{fig:netdistrib} shows the cumulative distribution of
servers as a function of the total number of ASes to which they belong.
We further separate the analysis between IPv4 and IPv6 servers, as
well as those servers that are active (score $\geq10$) or inactive.  We
find that the servers are concentrated in a small number of networks,
with 50\% of the servers belonging to fewer than 100 ASes.
Restricting the scope to just those servers that are active reduces
the overall ASN diversity.  Further, the addition of the IPv6 servers
does not add additional AS diversity, as evidenced by the intersection
of the IPv4 and IPv6 lines.   

%To better understand the types of individuals and organizations that volunteer
%their NTP servers to the \pool, 
We used IPinfo.io~\cite{ipinfo} to categorize
the \emph{types} of \acp{AS} that the \pool's servers are located within. Of the 15k
total servers our enumeration discovered (\S\ref{sec:data:scraper}), slightly
more than half (8,500, 54\%) were located in cloud hosting providers. Over 800
hosting companies are represented in this count, but the NTP server IPs in cloud
providers are disproportionately concentrated in only a few providers. Hetzner
has hosted 1,049 unique NTP IP addresses, or nearly 7\% of all \pool IP
addresses. Similarly, OVH hosts or has hosted nearly 5\% (732) of all \pool IPs;
Vultr, DigitalOcean, Akamai, Oracle and Amazon all contribute more than an
additional 1\% each as well.

An additional 6,036 (38\%) NTP server IPs were labeled with IPInfo's ``ISP''
category, which encompasses both large transit providers like Hurricane
Electric, but also customer \acp{AS} like Comcast, KPN, and Vodafone.  Of these,
Hurricane Electric is most common (397 \pool IPs), with Comcast (331) and
Deutsche Telekom (237) rounding out the top three.

The remaining 8\% of \pool server IPs belong to educational networks (476),
businesses (395) such as Alibaba, Apple, and Facebook, and governmental networks
(84) such as Hungary's KIFU Governmental Information Technology Development
Agency and the US National Institute of Standards and Technology (NIST).
One-hundred eighty-nine server IPs could not be categorized by IPInfo. 

These results indicate that \pool servers are being run predominantly in cloud
hosting networks. While the high availability of most cloud providers might be
viewed as a boon to the \pool's resilience, the fact that large swathes of
server IPs are found in only a small number of \acp{AS} indicates that the \pool
relies heavily on a few underlying providers. 

Finally, note that because most of the 15k \pool addresses are no longer active NTP
servers, it is possible that the IP addresses formerly in the \pool have since
been reassigned to a different \ac{AS}. This might confound our \ac{AS} type
analysis if the new \ac{AS} is of a different type. However, we believe that
this type of error is relatively uncommon and does not meaningfully affect the
overall distribution of NTP Pool server ASes. 

\subsection{IPv6 Servers}

Next, we examine the IPv6 servers within the pool, their connectivity,
and inferences we can make from the addresses.  First, we note that
one method for obtaining IPv6 connectivity in the absence of native
IPv6 is to utilize an IPv6-in-IPv4 tunnel, and a popular service
provider of this is Hurricane Electric's ``Tunnel Broker''~\cite{he}.
We identify Hurricane Electric tunnels via their registered IPv6
prefix of \texttt{2001:470::/32}.  A total of 337 servers are within
this prefix and likely connected via this tunnel broker.

Next, we use the \texttt{addr6} tool to characterize the interface
identifiers (IIDs) of the IPv6 server addresses.  \Cref{tab:iid} shows
the distribution of IIDs, across both active (score $\geq10$) and inactive
IPv6 servers.  Approximately 5\% of the servers use EUI-64 and embed
their interface's MAC address, while over 20\% have random, Privacy
Extension (PE) addresses.  Approximately 50\% of the addresses are
``low-byte,'' where the most significant bytes of the IID are zero
indicating that the address was likely manually configured.  

\begin{table}
\caption{Pool IPv6 Server Address IID Categorization: Low score
servers are more likely to use EUI64 and Privacy Extension (PE) addresses.}
\label{tab:iid}
\centering
\begin{tabular}{|l|r|r|r|}\hline
\rowcolor{black!30} \textbf{Type} & \textbf{All} & \textbf{score$\geq$10} &
    \textbf{score$<$10}\\\hline
Low-Byte   & 48.1\% & 50.7\% & 46.5\% \\\hline
Embed-IPv4 &  8.2\% &  8.5\% &  8.1\% \\\hline
Embed-port &  8.8\% & 12.0\% &  6.8\% \\\hline
EUI64      &  5.4\% &  3.1\% &  6.8\% \\\hline
PE         & 21.0\% & 16.7\% & 23.7\% \\\hline
Other      &  8.5\% &  9.0\% &  8.1\% \\\hline
\end{tabular}
\end{table}

PE and EUI-64 addresses are more typical of client hosts, whereas
low-byte and embedded are more typical of long-lived infrastructure
hosts.  Interestingly, the active servers are more likely to use a
low-byte address, while the inactive servers are more likely to use PE
and EUI-64. However, server AS-specific nuances are at play here, as well: for
instance, 169 of 205 (82\%) of the hosting provider Vultr's (AS20473) \vsix IP
addresses are \eui. On the other hand, 0 of the 89 Ionos (AS8560) \vsix IP
addresses are \eui.

\section{Pool Monopolization}
\label{sec:attack}

The canonical \pool definition of a ``server'' is an IP
address.  A single physical machine may have multiple network
interfaces, both physical and virtual, and may have multiple
IP addresses assigned to an interface.  Thus, there is a many-to-one
mapping of \pool server IP addresses to hosts.  For example, an 
NTP host may have a single interface with one IPv4 address and
two IPv6 addresses assigned to it -- if the operator of this host
registers all three addresses in the \pool, the \pool sees these
as three servers.

\subsection{Monitor Only Mode}
% SELECT s.Address, a.Org, a.Name, s.Zones FROM Meta m, Servers s,
% Accounts a WHERE Monitor=True AND Deleted=False AND m.Id=s.Id AND s.AcctId=a.AcctId;
% Qs:
%  - lifetime of these guise
%  - zone distrib
%  - Lots of HE tunnels1007
%  - big v6 clusters of adjacent addrs
% 
% SELECT COUNT(s.Address) FROM Meta m, Servers s WHERE Monitor=True AND Deleted=False AND Proto=6 AND m.Id=s.Id;
%
% wowow, 2,679 in "monitor ONLY"
%  1672 of these are v4
%  1007 of these are v6
%   115 of those are HE tunnels

Of note, one of the netspeed rates is zero, which corresponds to a
``monitor only'' mode.  In this mode, the server is a member of the
pool, but will not be included in any DNS responses (and, hence,
should not receive any client NTP queries as a result of the pool).
However, monitor only mode servers are queried by the pool monitors to
determine the quality of time they are providing. 

The IP addresses of the pool's monitoring infrastructure are not
published.  However, as described in prior works~\cite{kwon2023did}, monitor only
mode permits a server to readily determine the pool's current
monitors.  With knowledge of the monitors, a malicious server can
selectively respond, providing good time to the pool monitors, while
sending a different time to other clients, \eg as a part of a time skew
attack.

%9,955 / 5,725  = 15680

Our scraper finds all instances of servers with a netspeed of zero.
We find that 2,679 of the 15,680 servers (17.1\%) in our dataset are
operating in monitor only mode.  Among these, 1,672 are IPv4 servers
(62.4\%) while 1,007 are IPv6 servers (37.6\%).  While these monitor only mode
servers may be innocuous, adopting prior recommendations to use
ephemeral addresses for the monitors is necessary to defend against
attacks that rely on discovering the monitors.

\subsection{Residual Traffic}

%\todoer{What is the story here?  To keep the narrative consistent, 
%should we say that an adversary trying to snarf traffic need not even
%keep the server registered in the pool? Expand take-aways.}

\begin{figure}[t!]
        \centering
        \resizebox{1.0\columnwidth}{!}{\includegraphics{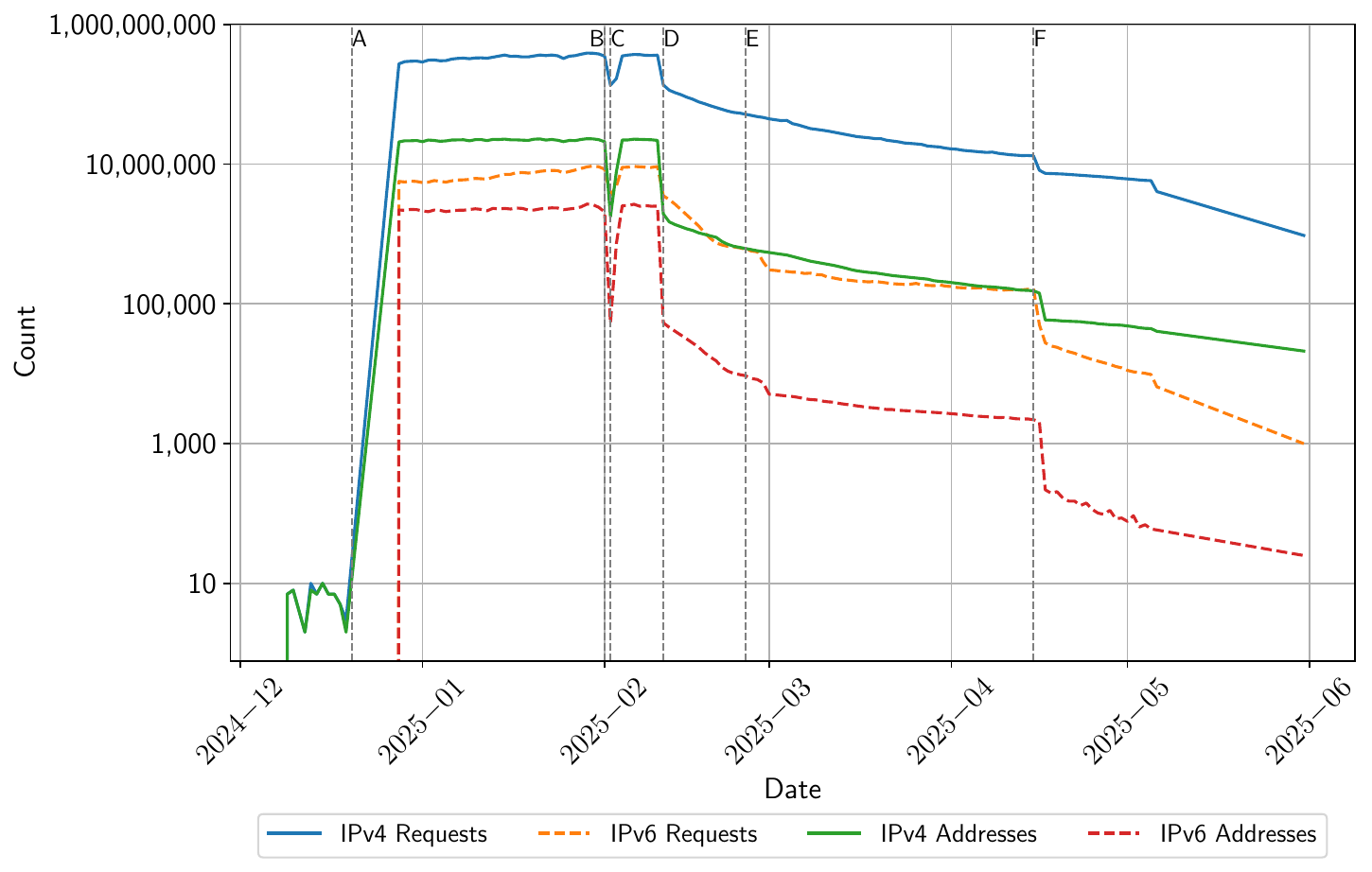}}
        \caption{Residual Traffic Experiment: Queries and unique
         addresses observed at a server per day after joining the pool (\textbf{A}), 
         scheduling for removal (\textbf{B}), canceling the removal
         (\textbf{C}), scheduling for removal again (\textbf{D}), removal
         (\textbf{E}), and termination of the NTP daemon (\textbf{F}).}
        \label{fig:erik_traffic}
\end{figure}

Our work demonstrates that an attacker need not even keep their server active in
the \pool to mount some attacks against NTP clients.
The \pool indicates that running an NTP server is a long-term commitment and
that traffic may persist long after a server's removal from the pool. To understand
the contours of this ``residual'' NTP traffic after a server has been removed
from the pool, we provisioned a new, dual-stacked NTP server in a US cloud
hosting provider. Figure~\ref{fig:erik_traffic} depicts the course of our
experiment, displaying the total number of \vfour/\vsix NTP requests and number
of unique \vfour/\vsix clients in per-day bins.
%\todoer{Explain what the bin sizes are for this plot.}

Before adding our NTP server to the \pool, it received some intermittent
NTP scanning on \vfour.  Several days after configuring our NTP server, we added
it to the \pool (time \textbf{A}). Our server quickly reached a steady state
rate of \vfour and \vsix NTP requests ($\sim$350M \vfour/7M \vsix requests per
day). At time \textbf{B}, we scheduled our server for removal from the \pool via
the \pool's web interface. Although the deletion date was set four days in the
future, it immediately began receiving fewer NTP requests. At time \textbf{C},
we canceled the impending deletion for our server. It soon returned to the
steady state rate of NTP requests. We again scheduled our NTP server for
deletion at time \textbf{D}, this time with a deletion date of two weeks in the
future (time \textbf{E}). At time \textbf{E}, our server was no longer
associated with the \pool system, but continued to receive a reduced steady
state number of NTP requests over the course of the next month ($\sim$25M
\vfour/200k \vsix). This is likely due to NTP clients that had obtained our
server's address continuing to keep it cached for unexpectedly long periods of
time.  Because our server still served accurate time, these clients were unaware
of the server's removal from the \pool. Finally, at time \textbf{F}, we stopped
our NTP daemon. This caused another significant decrease in the number of
requests and clients. 

%This experiment demonstrates that even if an attacker removes their server from the
%\pool, 
Our results show that many clients will cache an NTP server's IP address and
continue to use the server for time synchronization even months after removal
from the NTP Pool. Continuing to provide time suffices for hundreds of thousands
of clients to keep using our NTP server. This fact enables an attacker
interested in \eg skewing the time of their victims, to do so days or weeks
after they are no longer being monitored by the \pool's monitors.  

\subsection{Monopolization Attack}

Finally, we consider the power of an informed adversary to 
execute a monopolization attack.  Prior work also observed that
some zones are well populated, while others contain few servers -- as a
result, an adversarial node could join and potentially ``take over''
these zones~\cite{perry2021devil,moura2024deep}. In particular, Perry \etal
empirically determined the number of NTP servers an attacker would need to
contribute to five large pool zones (US, CA, UK, DE, and FR) to reach 50\% of
the total traffic for those zones~\cite{perry2021devil}. In these zones, they
discovered that an adversary need contribute 50-250 servers to reach the 50\%
threshold. In contrast, we use the computed netspeed values derived from the
``counts'' API endpoint (\S\ref{sec:data:netspeed}) to mathematically deduce the
number of servers required to monopolize the traffic of \emph{any} zone --
without needing to add servers to the zone \emph{a priori}. 

More precisely, for a given country or zone, let $n$ represent the current aggregate
netspeed as gathered via the \pool API and $m$ be the maximum possible
netspeed of any individual server; currently $m = 3000000$.  Assuming
an attack to capture a fraction $f$ of total NTP queries, then the
number of attack servers $S$ required is:
\begin{equation}
  S = \Big\lceil \frac{nf}{m(1-f)} \Big\rceil
\end{equation}

% vars:
%   f = fraction of traffic to capture
%   n = aggregate current netspeed
%   m = max netspeed
%   S = ceil(f * n / m)

% newnetspeed = oldnetspeed + (m*s)
% attacknetspeed = (m*s)
% fraccaptured = attacknetspeed / newnetspeed
%            f = (m*s) / (n + (m*s))
%            (n + (m*s)) * f = m*s
%            nf + fms = ms
%            ms - fms = nf
%            ms(1 - f) = nf
%             s = nf / m(1-f)
% While the large
% zones prior work studied require relatively many servers to monopolize the NTP
% traffic, the vast majority of zones require relatively few.

\begin{figure}[t!]
        \centering
        \resizebox{0.9\columnwidth}{!}{\includegraphics{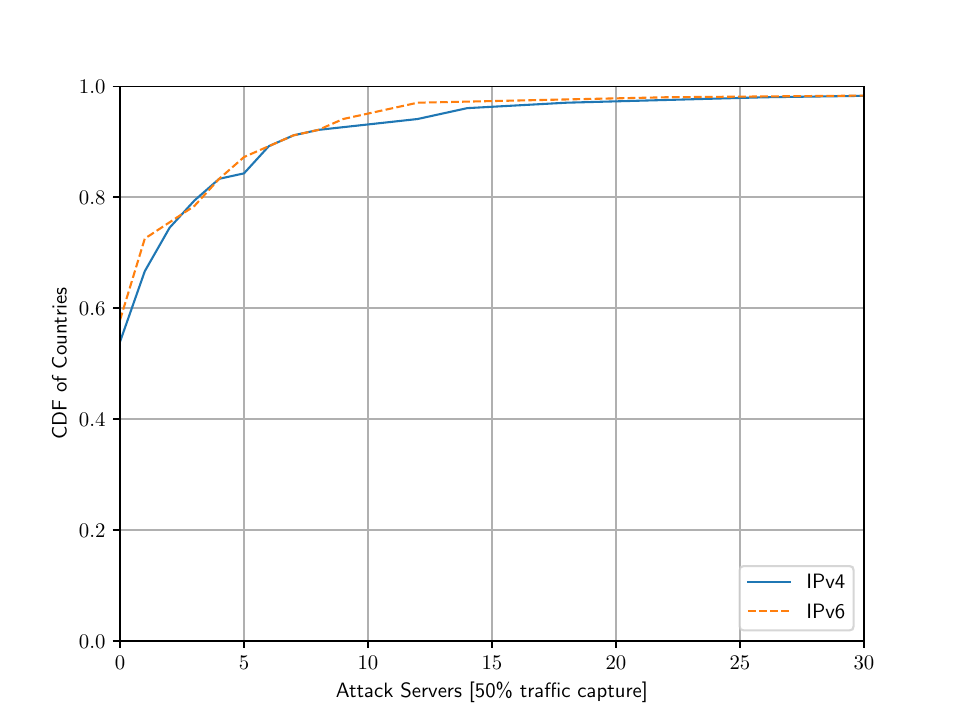}}
        \caption{Potential for traffic monopolization attack:
per-country number of required servers to capture at least half of the
zone's traffic.  90\% of countries require 10 or fewer attacking
servers to successfully execute the attack.}
        \label{fig:attackservers}
\end{figure}

% Our netspeed data provides a more nuanced view: an 
% adversary need not have the most servers in a zone, but instead, can
% tailor their attack to ensure that their netspeed is a significant
% or majority fraction of the aggregate netspeed in the zone.  
% 
We posit an adversary that wishes to receive at least half of the
traffic for a particular country, \ie contribute at least half of
the aggregate netspeed.  Further, we assume that the adversary sets
their netspeed to the maximum possible speed (3Gbps).
\Cref{fig:attackservers} displays, across all country zones, the 
number of servers the adversary
would require to capture half of the traffic.  More than half of the
countries would be compromised in this fashion by a single attacking
server, while the next 40\% of countries would require only 10 attack
servers.  

We conclude that the per-zone robustness to such attacks is
relatively low for 90\% of all countries.
Further, an attacker with the ability to create IPv6 aliases can
today effectively create an arbitrary number of servers to perform
the traffic monopolization attack on any country.  

\subsection{Monopolization Attack in Practice}

Last, we execute a limited version of the monopolization attack
in practice to demonstrate its feasibility in the wild, as well
as to validate our findings.  Note that while we ``attack'' the
zone, all of the servers we use in the experiment return good,
valid time -- hence, we did not disrupt the \pool, its clients,
or any host's notion of time.  Instead, we demonstrate the ability
to gain the preponderance of traffic within a zone via a capacity
informed adversary.

We performed our experiment on August 5, 2025.  We elected to target
the \texttt{.hu} zone (Hungary) as it contained six IPv6 active (score
$\geq$ 10) servers.  To understand our effect on clients within the zone,
we examine both the ``counts'' and the ``answers'' pool API endpoints
(\Cref{tab:endpoints}).  Recall that the ``answers'' endpoint 
reports the total count of times that each server was included in
DNS answers -- thereby allowing us to observe how the pool shifts
traffic.

As a baseline, prior to the attack, we observe that the zone has a
combined reported aggregate netspeed of 4.101Gbps.  Among the six
participating servers, four of the servers were apportioned for 24.4\%
of the DNS responses each ($\sim$71,000 answers per hour), one server
apportioned 2.4\% ($\sim$13,600 answers per hour) and one server apportioned
less than 0.1\% (134 responses per hour).

We then configured and added two IPv6 servers in the \texttt{.hu}
zone to mimic a monopoly attack.  Each attacking server 
was set to the maximum netspeed of 3Gbps.  After approximately
one day, we re-examined the distribution of netspeed and DNS answer
counts.  The two attack servers were each apportioned 29.7\% of the
netspeed and were included in $\sim$61,000 DNS answers per hour. The previous six
servers went down to 9.9\% for four ($\sim$36,000 answers per hour), 1.0\% for
one ($\sim$4,500 answers per hour), and almost 0\% (45 answers per hour) for the last.  By
DNS answer count, the
attack servers were each included in 23.4\% of the total DNS answers
within the period, for a total of 46.8\%.  Through manual
investigation, we find that this answer fraction was slightly lower than
expected as the two servers are also included in the global and
continent zones, thereby lowering their overall contribution to the
country zone.  However, this small experiment validates the ability
for a weak adversary to obtain a large fraction of the total country's
pool query traffic with only minimal resources (two servers
in-country). This problem is particularly acute in \vsix, as obtaining large
quantities of \vsix addresses to volunteer as NTP servers is trivial (some VPS
providers assign as much as /64 prefixes to individual servers).

\section{Conclusions}
 
In this work, we take a fresh look at the \pool\ -- a volunteer driven
and widely used NTP infrastructure, by gathering more complete and
rich data than previously possible.  By analyzing aliases, accounts,
and network connectivity, we find that only approximately 20\% of
the participating servers are truly independent, and that
the \pool is less robust than previously believed.  We then examine
monopoly attacks, wherein the adversary captures the preponderance of
NTP traffic in a particular country or region, and show that most 
zones in the pool are readily vulnerable to such attacks by a capacity
informed adversary.  

Our results suggest that the pool should consider
longevity and reputation, as well as server independence, 
in the scoring algorithm.  As a first step, the pool's backend
DNS server selection algorithm could be modified to consider the
account owner, protocol family, lifetime, and \ac{ASN} in its decision process.
We have shared our findings and recommendations with the
\pool operators.

%who are actively considering actions to improve the
%robustness of the system.

%post from ask:
% "One of my “wishlist items” for next I pass through or refactor the
% zone generation code is to make the system put limits on how many
% servers a user will be offered from the same server operator, the
% same ASN, etc (or another variation would, in well served areas,
% artificially limiting the “server speed” along the same buckets)."
% https://community.ntppool.org/t/why-is-cloudflare-in-the-pool/2455/8

\section{Ethics}
\label{sec:ethics}

Our measurements require continued periodic polling of the \pool
website, including both an API end-point and per-server status web
pages that we scrape.  
We reviewed our web scraping methodology with our institute’s digital librarian who provided several guiding principles: 1. scrape only public data; 2. ensure data is not covered by copyright; 3. obey any terms of service; 4. obey any rules provided to scrapers via robots.txt; 5. use the minimal query load possible; 6. use APIs when available; and 7. do not redistribute the data. 
We followed these principles and note that the \pool website is
publicly available, does not contain any personally identifiable
information or information on individuals and is not covered by copyright. 
Further, we follow the terms of
service and the robots.txt rules, and the API for data available via
an API. 

The website primarily
provides statistics and is decoupled from the operation of the pool's
time service -- \ie if the website were to fail, the pool's DNS and
NTP servers would continue to operate and provide time.  
We followed established best practices for ethical network measurements 
and minimized 
instantaneous load or any potential
service impact on the website by querying for new servers on average
once every 90 minutes and gathering updated statistics on existing
servers only once a day, where we wait an average of 5 seconds between
any two queries.   

Because our findings have potential security
implications for the pool, we do not make our complete dataset 
publicly available and
have shared our findings with the \pool project operators such that
they can make informed decisions on improving the project's
resilience. 

%\appendix
%\input{stratum}

\section*{Acknowledgment}

The authors would like to thank Peter Bartoli, Giovane Moura, Georgios Smaragdakis,
and the anonymous reviewers for invaluable constructive
feedback.  Thanks to Adam Shotland and Liam Carter for gathering and
providing the BigQuery data.

\newpage


\bibliographystyle{IEEEtran}
\bibliography{conferences,ntp}

\begin{acronym}
  \acro{AS}{Autonomous System}
  \acrodefplural{AS}[ASes]{Autonomous Systems}
  \acro{ASN}{\ac{AS} Number}
  \acro{CPE}{Customer Premises Equipment}
  \acro{EUI-64}{Extended Unique Identifier-64}
  \acro{MAC}{Media Access Control}
  \acro{NTP}{Network Time Protocol}
\end{acronym}

\end{document}